\pdfoutput=1
\def\arxivbuild{1}
\documentclass[preprint,12pt]{elsarticle}

\usepackage{amsmath,amssymb}
\usepackage{graphicx}
\usepackage{booktabs}
\usepackage{array}
\usepackage{xcolor}
\usepackage[colorlinks=true,linkcolor=blue,citecolor=blue,urlcolor=blue]{hyperref}
\usepackage{tikz}
\usepackage{float}
\floatstyle{plaintop}
\newfloat{algorithm}{tbp}{loa}
\floatname{algorithm}{Algorithm}

\newif\ifreaderq \readerqtrue
\ifdefined\arxivbuild \readerqfalse \fi
\ifreaderq
  
\else
  \usepackage{comment}
  \excludecomment{readerq}
\fi
\newcommand{\vpar}{v_\parallel}
\newcommand{\vperp}{v_\perp}
\newcommand{\dd}{\mathrm{d}}

\makeatletter
\def\ps@pprintTitle{\let\@oddhead\@empty\let\@evenhead\@empty
  \let\@oddfoot\@empty\let\@evenfoot\@empty}
\makeatother

\begin{document}

\begin{frontmatter}

\title{KINTORA: a semi-Lagrangian discontinuous Galerkin drift-kinetic,
unstructured mesh tokamak solver}

\author{Mark F.~Adams}
\ead{mfadams@lbl.gov}
\address{Applied Mathematics and Computational Research Division,
Lawrence Berkeley National Laboratory, Berkeley, CA, USA}

\begin{abstract}
Particle-in-cell codes support unstructured meshes but carry a
statistical noise floor; continuum gyrokinetic codes are free of that
noise and mostly use structured grids.  We present a continuum
drift-kinetic solver on unstructured triangular poloidal planes extruded
toroidally: a semi-Lagrangian discontinuous Galerkin discretization whose
parallel advance is a map along the exact field line in the
flux-coordinate independent manner, so the mesh is never aligned with the
field, with adiabatic electrons and a long-wavelength polarization field
solve.  Every operator
is verified against an independently computed reference, from Landau
damping to the geodesic-acoustic root and a linear
ion-temperature-gradient eigenvalue problem, every measured rate within
4\% except the geodesic-acoustic damping, within 12\%.  A linear
toroidal ion-temperature-gradient demonstration on a
canonical-Maxwellian equilibrium recovers a radially coherent mode
whose growth rate, $0.065\ c_s/a$, is stationary in time.  The
collisionless zonal-flow residual rises with poloidal refinement from
$0.72$ to $0.87$ of the model's own orbit-quadrature prediction for the
measured observable; an Eulerian discretization of the same model stops
moving with the mesh at $0.82$ and rises $16\%$ when the parallel velocity
resolution is doubled.  That resolution and the prediction's radial
locality, from which global gyrokinetic codes also depart at finite
$\rho^*$, are both candidates for the remaining gap.
The solver is fully device resident, and every
gate of the verification ladder is reproduced between CPU and GPU builds
and across toroidal process counts.  The code, the data behind every table and an artifact
description are in a public repository.
\end{abstract}


\end{frontmatter}

\section{Introduction}
\label{sec:intro}

Gyrokinetic codes for tokamak core turbulence differ in two ways, in how
they represent phase space and in the mesh they use.  Continuum codes,
Eulerian or semi-Lagrangian, such as GENE~\cite{Jenko2000},
GS2~\cite{Kotschenreuther1995}, CGYRO~\cite{Candy2016} and
GYSELA~\cite{Grandgirard2006,Grandgirard2016}, resolve the distribution
function on a grid, with no sampling noise, and these four use structured
grids on nested flux surfaces.  Continuum codes for the edge, such as
Gkeyll and GENE-X, reach open field lines and the X-point.  Gkeyll's discontinuous Galerkin (DG) scheme
runs full-$f$ gyrokinetics on open field lines with sheath
boundaries~\cite{Shi2017,Mandell2020,Hoffmann2026NF}, and its
field-aligned coordinates have been constructed for X-point
geometry~\cite{Shukla2026}.  GENE-X replaces flux-aligned coordinates with the flux-coordinate
independent (FCI) approach~\cite{Michels2021}.  Particle-in-cell
codes such as XGC~\cite{Ku2009} handle unstructured meshes and open
field lines at the cost of a sampling noise floor, which linear and
near-marginal problems must filter out.  Jo et al.\ discretize the
distribution function on an \emph{unstructured} mesh with an explicit
DG solver~\cite{Jo2025}, on prisms that extend unstructured poloidal
triangles along the field; that solver does not yet include a field
solve.  The FCI parallel advance of the fluid edge codes GRILLIX and
BOUT++~\cite{Hariri2013,Stegmeir2016} is another route to a continuum
code on an unstructured mesh: it traces the field line exactly and
interpolates the plane at its foot, so the mesh need not be aligned with
the field (Section~\ref{sec:geometry}).

This paper presents such a solver, KINTORA, a semi-Lagrangian
discontinuous Galerkin (SLDG) discretization of the drift-kinetic
equation on unstructured triangular poloidal planes extruded toroidally.
The semi-Lagrangian Vlasov method goes back to Cheng and
Knorr~\cite{ChengKnorr1976} and Sonnendr\"ucker et
al.~\cite{Sonnendrucker1999}.  The discontinuous Galerkin variant, which
conserves mass for a shift along a grid axis, follows Crouseilles,
Mehrenberger and collaborators~\cite{CrouseillesMehrenbergerVecil2011}
and Einkemmer~\cite{Einkemmer2016}.  Kormann, Reuter and Rampp decompose
a six-dimensional semi-Lagrangian Vlasov--Poisson solver over all of its
dimensions~\cite{Kormann2019}, and BSL6D runs that algorithm on CPU, AMD
and NVIDIA nodes from one Kokkos code base~\cite{Schild2024}.

Of the codes above, GYSELA is the one that is semi-Lagrangian; it was
drift-kinetic in its original form, and validated on slab ITG against the
analytic dispersion relation~\cite{Grandgirard2006}.  We follow its
validation template where it applies, but not its discretization.  Its
5D advective form sets two divergence source terms to zero, and its
authors report charge and toroidal momentum conserved to about $10\%$,
and energy not conserved to that level~\cite{Grandgirard2016}.  Gkeyll's
DG scheme conserves particles exactly and energy in the semi-discrete
form~\cite{Mandell2020}.  In the present scheme
the transfer along each field line and the shift in parallel velocity
conserve mass under the $L^2$ projection, and the evaluations within the
poloidal plane do not; Section~\ref{sec:ladder-cons} measures the
resulting defects.

The model has adiabatic electrons and drift-kinetic ions, with the
gyroaverage $J_0 \to 1$ in both the Vlasov and the field equation.  Two
equilibria carry the data.  The toroidal rungs of the verification ladder
and the rates of Sections~\ref{sec:cbc} and~\ref{sec:rh} are measured on
the circular concentric one; three further rungs run with uniform $B$.
The field path, the field-line trace, the drift kinematics
(Section~\ref{sec:shaped}) and one geodesic-acoustic frequency
(Section~\ref{sec:gam}) are measured on the shaped Solov'ev equilibrium
of Figure~\ref{fig:algorithm}.  The X-point motivates the design, and
the field-line backtrace sits behind an interface that does not assume
nested surfaces.

This paper's contributions are:
\begin{itemize}
\item A continuum drift-kinetic solver on unstructured poloidal planes, with
  a field-line-map parallel advance (Sections~\ref{sec:method}
  and~\ref{sec:geometry}).
\item A verification ladder in one table: each operator against an
  independently computed reference, on CPU and GPU (Section~\ref{sec:ladder}).
\item Conservation under joint refinement: the energy drift is bounded; the
  mass defect grows in time and does not fall with refinement
  (Section~\ref{sec:ladder-cons}).
\item A linear toroidal ITG mode on a canonical-Maxwellian equilibrium,
  growing at a stationary $0.065\,c_s/a$ (Section~\ref{sec:cbc-demo}).
\item A closed-form map from label to effective gradients; at
  $\rho^* = 0.02$ this equilibrium class contains no CBC point
  (Section~\ref{sec:cbc-dilution}).
\item The geodesic acoustic frequency within $0.6\%$ of the model's own
  continuum, and its damping within $12\%$ (Section~\ref{sec:gam}).
\item Two discretizations put the Rosenbluth--Hinton residual at
  $0.82$--$0.87$ of the model's radially local prediction
  (Section~\ref{sec:residual}).
\end{itemize}
Section~\ref{sec:performance} gives the cost; the parallel and GPU design
is the subject of a companion paper~\cite{AdamsHPC2026}.  The appendices
state the verification methodology and derive the geodesic-acoustic root
and the model's neoclassical polarization.

\section{Model and discretization}
\label{sec:method}

\subsection{The drift-kinetic model}

We evolve the full ion distribution $f(\mathbf{x}, \vpar, \mu, t)$ with the
magnetic moment $\mu$ an adiabatic invariant:
\begin{equation}
\frac{\partial f}{\partial t}
+ \left( \vpar \mathbf{b} + \mathbf{v}_d + \mathbf{v}_E \right) \cdot
  \nabla f \big|_{\mu}
+ \left( E_\parallel - \mu\, \nabla_\parallel B
  + \vpar\, \mathbf{v}_E \cdot \nabla \ln B \right)
  \frac{\partial f}{\partial \vpar} = 0 ,
\label{eq:dk}
\end{equation}
in units with unit charge and mass, where $\mathbf{b}$ is the field
direction, $E_\parallel = -\nabla_\parallel \phi$,
\begin{equation}
\mathbf{v}_d = \frac{\vpar^2 + \mu B}{B}\, \mathbf{b} \times \nabla \ln B
\end{equation}
is the combined grad-$B$ and curvature drift of a vacuum field, and
$\mathbf{v}_E = (\mathbf{b} \times \nabla_\perp \phi)/B$.  The last term,
the parallel acceleration $\vpar\, \mathbf{v}_E \cdot \nabla \ln B$, is the
electric field's work on the curvature drift
$\mathbf{v}_c = (\vpar^2/B)\, \mathbf{b} \times \nabla \ln B$: it equals
$-(\mathbf{v}_c \cdot \nabla \phi)/\vpar$, so it changes $\vpar^2/2$ at
the rate $-\mathbf{v}_c \cdot \nabla \phi$.  It is the
$\vpar \nabla \times \mathbf{b}$ part of the modified field
$\mathbf{B}^* = \mathbf{B} + \vpar \nabla \times \mathbf{b}$ of the
guiding-centre equations~\cite{BrizardHahm2007}.  Its $\vpar$-derivative
is minus the phase-space divergence of $\mathbf{v}_E$,
$B^{-1} \nabla \cdot (B\, \mathbf{v}_E)$, so the E$\times$B flow and this
acceleration together conserve phase-space volume.
Section~\ref{sec:rh} applies it as a dilation of $\vpar$.

The field is the long-wavelength limit of the gyrokinetic
quasi-neutrality condition with adiabatic
electrons~\cite{Lee1983,Dubin1983,Hahm1988},
\begin{equation*}
-\nabla \cdot \left( \frac{m_i n_0}{e B^2}\, \nabla_\perp \phi \right)
+ \frac{e n_0}{T_e} \left( \phi - \langle \phi \rangle \right) = \delta n_i ,
\end{equation*}
in which the ion polarization density and the electron response balance
the perturbation $\delta n_i$ of the ion guiding-centre density, and
$\langle\phi\rangle$ is the flux-surface average, to which the electrons
do not respond.  In the units of Equation~(\ref{eq:dk}), with $\phi$ in
units of $T_i/e$ and $\delta n = \delta n_i/n_0$, multiplying by $T_e/e$
gives the equation solved per poloidal plane on the nodal space of the
state,
\begin{equation}
-\nabla \cdot \left( n_0(\mathbf{x})\, \rho_s^2\, \nabla_\perp \phi \right)
+ n_0(\mathbf{x}) \left( \phi - \langle \phi \rangle \right)
= n_0(\mathbf{x})\, \tau\, \delta n ,
\qquad \rho_s^2 = \tau / B^2 ,
\label{eq:field}
\end{equation}
with $\tau = T_e/T_i$ and natural boundary conditions.  The toroidal ITG
and zonal-flow runs of Sections~\ref{sec:cbc} and~\ref{sec:rh} carry the
average, and Section~\ref{sec:rh} gives its discretization; without it
the operator is symmetric positive definite.  The discretization keeps
$n_0(\mathbf{x})$ inside the divergence in the toroidal measure: the
weight at node $p$ is $n_{0p}w_p$, with $n_{0p}$ the grid's own moment of
$f_0$, which carries the Jacobian $R$.  The transported stages need no
such factor, because their velocity measure omits $2\pi B$, their spatial
measure omits $R$, and $BR = B_0R_0$ is constant (\ref{app:residual}).
$\rho_s^2$ uses the electron temperature and the reference ion
temperature, and is not profiled.

The code carries three field models.  The concentric one has
$B = B_0 R_0 / R$ on circular flux surfaces, and its field lines wind with
a sheared safety factor $q(r) = q_0 + (q_a - q_0)(r/a)^2$, which enters the
field-line maps, the parallel gradient and the mirror force through
$\iota = 1/q$ (the CBC configuration of Section~\ref{sec:cbc} has
$q(r_0) = 1.4$ and $\hat{s} = 0.78$).  The poloidal field's
$O\!\left((r/qR_0)^2\right)$ contribution to $|B|$ is neglected, but the
line-length pitch factor $\sqrt{1 + (r\iota/R_0)^2}$ is kept in the
parallel streaming rate.  The large-aspect-ratio model
$B = B_0(1 - \epsilon\cos\theta)$ makes the along-line dynamics exactly
the pendulum, so elliptic-integral bounce theory is its exact continuum
reference (Section~\ref{sec:ladder-bounce}).  The third is the analytic
Solov'ev equilibrium of Figure~\ref{fig:algorithm}, whose field line is
traced in its $\psi(R,Z)$ and whose $|B|$ varies in both poloidal
coordinates (Section~\ref{sec:shaped}).

\subsection{Phase-space measure and conservation}
\label{sec:measure}

The velocity measure is $\dd^3 v = 2\pi B\, \dd\mu\, \dd\vpar$.  The
stream moves content along field lines, and a flux tube's cross-section
scales as $1/B$, so the stream conserves $f\, \vperp / B$ per unit
$(\vperp, \vpar)$; equivalently, the flux-tube measure $\dd\mu\,\dd\vpar$
per unit length is independent of $B$.  The discretization's moment
functionals use this measure exactly, so each stage's change of the mass
is measured in it (Section~\ref{sec:ladder-cons}).

One property of the concentric equilibrium sets the drift stage's design
there.  Both circular field models depend on the poloidal coordinate
$x = R - R_0$ alone, so $\mathbf{v}_d$ is vertical with magnitude
$(\vpar^2 + \mu B) D(x)$, $D = -(\dd B/\dd x)/B^2$.  A vertical field whose
magnitude depends only on $x$ is divergence-free, and so is its product
with any function of $x$, such as $1/B^k$:
\begin{equation}
\nabla \cdot \left( \mathbf{v}_d / B^k \right) = 0
\quad \text{for every } k .
\end{equation}
The drift is therefore incompressible in the flux-tube measure, and the
stage moves each node along its vertical characteristic to an exact foot,
since the displacement is constant along its own ray.  It preserves $\mu$
and leaves $\vperp$ unchanged, because the drift moves along contours of
$B$.  Its mass defect is its pointwise sampling error plus the
perturbation's flux through the open wall
(Sections~\ref{sec:ladder-drift} and~\ref{sec:ladder-cons}).  On the
shaped equilibrium the drift is the general guiding-centre displacement,
a vector per node and velocity point, and the stage evaluates the state at
that foot (Section~\ref{sec:shaped}).

\subsection{Discretization and splitting}
\label{sec:splitting}

The poloidal plane carries an unstructured triangular mesh with nodal
Lagrange elements ($P_2$ or $P_3$), and \emph{cell} means one of its
triangles.  The toroidal direction has $N_\phi$ DG elements, the
\emph{wedges}.  Each carries the Gauss--Lobatto--Legendre (GLL)
\emph{stations} of degree $p_\phi$, copies of the poloidal plane: three
per wedge at the default $p_\phi = 2$, with the end stations of
neighbouring wedges coincident in $\varphi$ but distinct data.  Velocity
space is a tensor grid: DG in $\vpar$ ($N_v$ \emph{segments},
one-dimensional elements with GLL nodes of degree $p_v$) by $N_w$
\emph{$\mu$ rows}, lines of constant magnetic moment.  The $\mu$ rows sit
at the GLL positions of equal intervals in $w = \sqrt{2B_0\mu}$ (eight
intervals of degree 2 in production, so $N_w = 24$), and the
perpendicular speed at a node is $w\sqrt{B(\mathbf{x})/B_0}$.  A
\emph{channel} is a ($\vpar$ node, $\mu$ row) pair, and a \emph{pencil} is
a channel at one poloidal node, the line of points along $\varphi$ that
the stream shifts.  Because $\mu$ is invariant along a field line and
under the drift, no stage moves content between $\mu$ rows: every stage is
a per-$\mu$-row operation (Figure~\ref{fig:velocity}), and the E$\times$B
term's in-plane gradient is taken at fixed $\mu$, as the model requires.

Figures~\ref{fig:algorithm} to~\ref{fig:velocity} draw one time step
and Algorithm~\ref{alg:step} lists it, with the same names.  The step is
Strang splitting: half-steps of the parallel stream and the drift, full
steps of the field solve, the velocity kick and the E$\times$B term, then
the second half-drift and the second half-stream.  Three placements put
the field-coupled quantities at the step's midpoint.  The drift is split
about the field solve, so the density the field is solved from is at the
midpoint of the drift's motion.  The E$\times$B term moves density by
$-(\Delta t/2)\,\nabla\cdot(n\,\mathbf{v}_E)$ over the half step, and the
predictor of step 2 adds that change; the kick moves no density.  The
kick's curvature exponent is taken at the midpoint of the E$\times$B
trajectory (step 5).  Without the predictor the $O(\Delta t)$ error in
$\phi$ feeds both drive terms and the step is first order; with it, the
toroidal ITG configuration of Section~\ref{sec:cbc} measured order 2.6.
Without these three placements the geodesic acoustic damping of
Section~\ref{sec:gam} is first order in the step.

\begin{figure}[t!]
\centering
\input{fig_algorithm_body}
\caption{One time step in configuration space, on the torus; the circled
numbers are the steps of Algorithm~\ref{alg:step}.  The eight planes are
the boundaries of the wedges, the toroidal DG elements.  The target plane carries the
coarsest flux-surface-aligned mesh of the up-down asymmetric Solov'ev
equilibrium of Section~\ref{sec:shaped}.  The field line through node $p$ is
drawn for one toroidal transit (thin dashed red): it lies in a flux
surface and its poloidal angle advances by $\iota\,\Delta\varphi$.  The
stream traces it back from $p$ (dashed) to its foot two wedges back;
$\times$ marks the plane crossings; the solid line is wedge $c$ shifted
by the same amount, the range of the overlap integral of
Figure~\ref{fig:strip}.  The shift is drawn at 1.25 wedges, several
times the production value.  At the far left, the drift stage: the
vertical fiber through a node and its foot inside a triangle.}
\label{fig:algorithm}
\end{figure}

\begin{figure}[t!]
\centering
\input{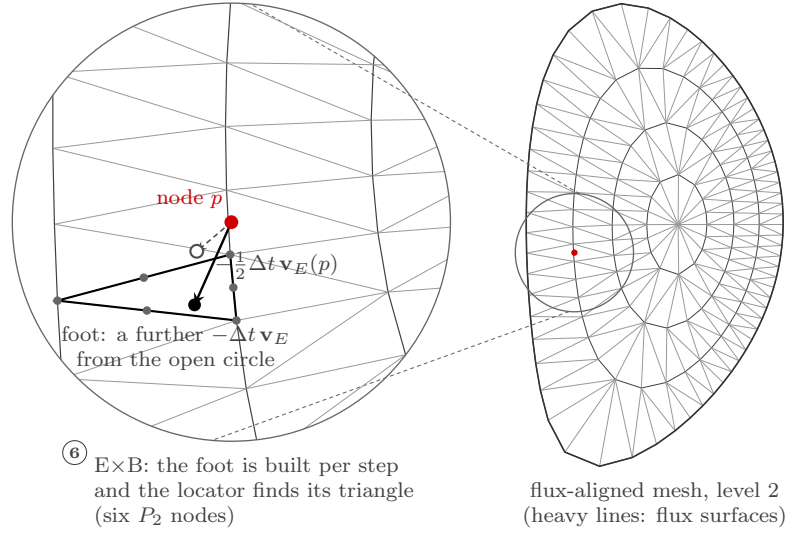}
\caption{The poloidal plane and the E$\times$B stage (step 6 of
Algorithm~\ref{alg:step}).  Right: the flux-surface-aligned mesh of
Figure~\ref{fig:algorithm} one level finer, face-on; the heavy lines are
the flux surfaces.  Left: the patch circled on the plane, magnified: the
half step from node $p$ to the open circle, the full step from the
velocity found there to the foot (Equation~\ref{eq:exbfoot}), and the six
$P_2$ nodes of the triangle the locator returns.  The displacement is
drawn at about one cell; the largest the demonstration of
Section~\ref{sec:cbc-demo} reaches is $0.30$ of the smallest cell edge.}
\label{fig:plane}
\end{figure}

\begin{figure}[t!]
\centering
\begin{tikzpicture}[font=\footnotesize, >=stealth, mesh/.style={black!45, line width=0.2pt}, bd/.style={black!80, line width=0.6pt}, tri/.style={black, line width=0.7pt}, fl/.style={red!80!black, line width=1.0pt}, fltr/.style={red!80!black, line width=0.6pt, dash pattern=on 1.9pt off 1.4pt}, lab/.style={font=\scriptsize, black!75, align=left, inner sep=1.5pt}, sn/.style={circle, draw=black!75, inner sep=0.4pt, minimum size=9pt, font=\tiny\bfseries, text=black!75}]
  \fill[black!25] (0.775,-0.080) rectangle (3.275,0.540);
  \draw[black!70, line width=0.5pt] (0.775,-0.080) rectangle (3.275,0.540);
  \draw[bd] (0.275,0.000) -- (6.600,0.000) (0.275,0.460) -- (6.600,0.460);
  \draw[bd] (1.400,0.000) -- (1.400,0.460);
  \draw[bd] (3.900,0.000) -- (3.900,0.460);
  \draw[bd] (6.400,0.000) -- (6.400,0.460);
  \node[lab, inner sep=0pt] at (0.055,0.230) {$\cdots$};
  \node[lab, inner sep=0pt] at (6.820,0.230) {$\cdots$};
  \fill[black] (1.400,0.230) circle (1.2pt);
  \fill[black] (2.650,0.230) circle (1.2pt);
  \fill[black] (3.900,0.230) circle (1.2pt);
  \fill[black] (3.900,0.230) circle (1.2pt);
  \fill[black] (5.150,0.230) circle (1.2pt);
  \fill[black] (6.400,0.230) circle (1.2pt);
  \node[lab, anchor=north] at (1.500,-0.020) {$c\!-\!1$};
  \node[lab, anchor=north] at (4.000,-0.020) {$c\!:\,0$};
  \node[lab, anchor=north] at (5.150,-0.020) {1};
  \node[lab, anchor=north] at (6.400,-0.020) {2};
  \draw[black!70, line width=0.5pt] (0.315,1.010) -- (3.860,1.010);
  \draw[black!70, line width=0.5pt] (0.315,0.940) -- (0.315,1.080) (3.860,0.940) -- (3.860,1.080);
  \node[lab, anchor=south] at (2.087,1.030) {rank 0};
  \draw[black!70, line width=0.5pt] (3.940,1.010) -- (6.560,1.010);
  \draw[black!70, line width=0.5pt] (3.940,0.940) -- (3.940,1.080) (6.560,0.940) -- (6.560,1.080);
  \node[lab, anchor=south] at (5.250,1.030) {rank 1};
  \draw[<->, black, line width=0.6pt] (3.400,1.460) -- (4.400,1.460);
  \node[lab, anchor=west] at (4.500,1.460) {ghost exchange: one wedge each way};
  \draw[red!80!black, line width=0.6pt, dash pattern=on 2pt off 1.5pt] (3.900,-0.080) rectangle (6.400,0.540);
  \fill[red!80!black] (6.400,0.230) circle (1.8pt);
  \fill[black] (3.275,0.230) circle (1.7pt);
  \node[lab, text=red!80!black, anchor=south] at (6.400,0.560) {node $p$};
  \node[lab, anchor=south] at (3.275,0.560) {foot};
  \draw[->, red!80!black, line width=0.8pt] (6.400,-0.450) -- (3.275,-0.450);
  \node[lab, anchor=north west, text=red!80!black] (L1) at (5.288,-0.550) {wedge $c$ (dashed), shifted by $v_\parallel g(r)\,\Delta t/2$ (shaded):\\ the transfer integrates the overlap with wedges $c\!-\!1$ and $c$};
  \node[sn, anchor=east, xshift=-1pt] (L2) at (L1.west) {8};
  \node[sn, anchor=east, xshift=-1pt] (L3) at (L2.west) {1};
  \node[lab, anchor=west] at (7.020,0.230) {$\varphi$};
  \node[lab, anchor=west, ] (L4) at (7.850,0.230) {filter: keep $|n| \le n_{\max}$ along $\varphi$};
  \node[sn, anchor=east, xshift=-1pt] (L5) at (L4.west) {4};
\end{tikzpicture}
\caption{The parallel semi-Lagrangian transfer (steps 1 and 8 of
Algorithm~\ref{alg:step}) on the toroidal grid: two wedges of the strip
along $\varphi$ with their three GLL stations each, the rank that owns
each wedge, and the one-wedge ghost exchange between neighbours.  Red is
node $p$ of Figure~\ref{fig:algorithm} at station $(c,2)$, the dashed red
box its wedge, black its foot, the shaded box that wedge shifted by
$\vpar g(r)\,\Delta t/2$; the transfer is the overlap integral of the
shifted wedge with the source wedges it covers.  The filter (step 4) acts
along the same strip.}
\label{fig:strip}
\end{figure}
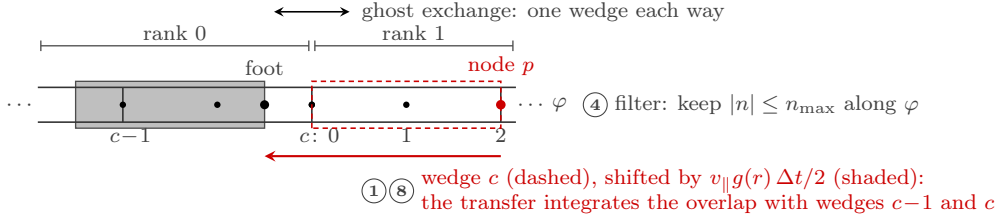

\begin{figure}[t!]
\centering
\begin{tikzpicture}[font=\footnotesize, >=stealth, mesh/.style={black!45, line width=0.2pt}, bd/.style={black!80, line width=0.6pt}, tri/.style={black, line width=0.7pt}, fl/.style={red!80!black, line width=1.0pt}, fltr/.style={red!80!black, line width=0.6pt, dash pattern=on 1.9pt off 1.4pt}, lab/.style={font=\scriptsize, black!75, align=left, inner sep=1.5pt}, sn/.style={circle, draw=black!75, inner sep=0.4pt, minimum size=9pt, font=\tiny\bfseries, text=black!75}]
  \fill[black!10] (1.500,0.680) rectangle (10.500,0.880);
  \fill[black!10] (1.500,1.460) rectangle (10.500,1.660);
  \fill[black!10] (1.500,2.240) rectangle (10.500,2.440);
  \fill[black!25] (5.055,0.660) rectangle (7.305,0.900);
  \draw[black!70, line width=0.5pt] (5.055,0.660) rectangle (7.305,0.900);
  \fill[black!25] (3.975,1.440) rectangle (6.225,1.680);
  \draw[black!70, line width=0.5pt] (3.975,1.440) rectangle (6.225,1.680);
  \fill[black!25] (2.625,2.220) rectangle (4.875,2.460);
  \draw[black!70, line width=0.5pt] (2.625,2.220) rectangle (4.875,2.460);
  \draw[black!60, line width=0.4pt] (1.500,0.000) -- (1.500,2.740) -- (1.425,2.810) -- (1.575,2.910) -- (1.425,2.990) -- (1.500,3.040);
  \draw[black!60, line width=0.4pt] (3.750,0.000) -- (3.750,2.740) -- (3.675,2.810) -- (3.825,2.910) -- (3.675,2.990) -- (3.750,3.040);
  \draw[black!60, line width=0.4pt] (6.000,0.000) -- (6.000,2.740) -- (5.925,2.810) -- (6.075,2.910) -- (5.925,2.990) -- (6.000,3.040);
  \draw[black!60, line width=0.4pt] (8.250,0.000) -- (8.250,2.740) -- (8.175,2.810) -- (8.325,2.910) -- (8.175,2.990) -- (8.250,3.040);
  \draw[black!60, line width=0.4pt] (10.500,0.000) -- (10.500,2.740) -- (10.425,2.810) -- (10.575,2.910) -- (10.425,2.990) -- (10.500,3.040);
  \draw[black!60, line width=0.4pt] (1.500,0.000) -- (10.500,0.000);
  \draw[black!60, line width=0.4pt] (1.500,1.560) -- (10.500,1.560);
  \draw[black!35, line width=0.3pt, dotted] (1.500,0.000) -- (10.500,0.000);
  \fill[black] (1.500,0.000) circle (1.0pt);
  \fill[black] (2.122,0.000) circle (1.0pt);
  \fill[black] (3.128,0.000) circle (1.0pt);
  \fill[black] (3.750,0.000) circle (1.0pt);
  \fill[black] (4.372,0.000) circle (1.0pt);
  \fill[black] (5.378,0.000) circle (1.0pt);
  \fill[black] (6.000,0.000) circle (1.0pt);
  \fill[black] (6.622,0.000) circle (1.0pt);
  \fill[black] (7.628,0.000) circle (1.0pt);
  \fill[black] (8.250,0.000) circle (1.0pt);
  \fill[black] (8.872,0.000) circle (1.0pt);
  \fill[black] (9.878,0.000) circle (1.0pt);
  \fill[black] (10.500,0.000) circle (1.0pt);
  \draw[black!35, line width=0.3pt, dotted] (1.500,0.780) -- (10.500,0.780);
  \fill[black] (1.500,0.780) circle (1.0pt);
  \fill[black] (2.122,0.780) circle (1.0pt);
  \fill[black] (3.128,0.780) circle (1.0pt);
  \fill[black] (3.750,0.780) circle (1.0pt);
  \fill[black] (4.372,0.780) circle (1.0pt);
  \fill[black] (5.378,0.780) circle (1.0pt);
  \fill[black] (6.000,0.780) circle (1.0pt);
  \fill[black] (6.622,0.780) circle (1.0pt);
  \fill[black] (7.628,0.780) circle (1.0pt);
  \fill[black] (8.250,0.780) circle (1.0pt);
  \fill[black] (8.872,0.780) circle (1.0pt);
  \fill[black] (9.878,0.780) circle (1.0pt);
  \fill[black] (10.500,0.780) circle (1.0pt);
  \draw[black!35, line width=0.3pt, dotted] (1.500,1.560) -- (10.500,1.560);
  \fill[black] (1.500,1.560) circle (1.0pt);
  \fill[black] (2.122,1.560) circle (1.0pt);
  \fill[black] (3.128,1.560) circle (1.0pt);
  \fill[black] (3.750,1.560) circle (1.0pt);
  \fill[black] (4.372,1.560) circle (1.0pt);
  \fill[black] (5.378,1.560) circle (1.0pt);
  \fill[black] (6.000,1.560) circle (1.0pt);
  \fill[black] (6.622,1.560) circle (1.0pt);
  \fill[black] (7.628,1.560) circle (1.0pt);
  \fill[black] (8.250,1.560) circle (1.0pt);
  \fill[black] (8.872,1.560) circle (1.0pt);
  \fill[black] (9.878,1.560) circle (1.0pt);
  \fill[black] (10.500,1.560) circle (1.0pt);
  \draw[black!35, line width=0.3pt, dotted] (1.500,2.340) -- (10.500,2.340);
  \fill[black] (1.500,2.340) circle (1.0pt);
  \fill[black] (2.122,2.340) circle (1.0pt);
  \fill[black] (3.128,2.340) circle (1.0pt);
  \fill[black] (3.750,2.340) circle (1.0pt);
  \fill[black] (4.372,2.340) circle (1.0pt);
  \fill[black] (5.378,2.340) circle (1.0pt);
  \fill[black] (6.000,2.340) circle (1.0pt);
  \fill[black] (6.622,2.340) circle (1.0pt);
  \fill[black] (7.628,2.340) circle (1.0pt);
  \fill[black] (8.250,2.340) circle (1.0pt);
  \fill[black] (8.872,2.340) circle (1.0pt);
  \fill[black] (9.878,2.340) circle (1.0pt);
  \fill[black] (10.500,2.340) circle (1.0pt);
  \draw[red!80!black, line width=0.6pt, dash pattern=on 2pt off 1.5pt] (6.000,0.590) rectangle (8.250,0.970);
  \fill[red!80!black] (6.622,0.780) circle (1.9pt);
  \fill[black] (5.677,0.780) circle (1.8pt);
  \draw[->, red!80!black, line width=0.8pt] (6.622,1.080) -- (5.677,1.080);
  \node[lab, text=red!80!black, anchor=west] at (6.702,0.460) {node};
  \node[lab, anchor=east] at (5.597,0.460) {foot};
  \draw[red!80!black, line width=0.6pt, dash pattern=on 2pt off 1.5pt] (6.000,1.370) rectangle (8.250,1.750);
  \fill[red!80!black] (7.628,1.560) circle (1.9pt);
  \fill[black] (5.603,1.560) circle (1.8pt);
  \draw[->, red!80!black, line width=0.8pt] (7.628,1.860) -- (5.603,1.860);
  \node[lab, text=red!80!black, anchor=west] at (7.708,1.240) {node};
  \node[lab, anchor=east] at (5.523,1.240) {foot};
  \draw[red!80!black, line width=0.6pt, dash pattern=on 2pt off 1.5pt] (6.000,2.150) rectangle (8.250,2.530);
  \fill[red!80!black] (8.250,2.340) circle (1.9pt);
  \fill[black] (4.875,2.340) circle (1.8pt);
  \draw[->, red!80!black, line width=0.8pt] (8.250,2.640) -- (4.875,2.640);
  \node[lab, text=red!80!black, anchor=west] at (8.330,2.020) {node};
  \node[lab, anchor=east] at (4.795,2.020) {foot};
  \node[lab, anchor=south west, text=red!80!black] (L1) at (1.950,3.140) {kick: a row shifted by $-(E_\parallel - \mu\,\partial_s B)\,\Delta t$, node $\to$ foot; rows never mix};
  \node[sn, anchor=east, xshift=-1pt] (L2) at (L1.west) {5};
  \draw[->, black] (1.350,-0.450) -- (10.700,-0.450);
  \node[lab, anchor=north] at (1.500,-0.500) {$-v_{\max}$};
  \node[lab, anchor=north] at (10.500,-0.500) {$+v_{\max}$};
  \draw[->, black] (1.050,-0.150) -- (1.050,3.290);
  \node[lab, anchor=east] at (1.000,0.000) {$w = 0$};
  \node[lab, anchor=south, rotate=90] at (0.250,1.520) {$\mu$ rows, $w = \sqrt{2 B_0 \mu}$};
  \node[lab, anchor=north, align=center] at (6.000,-0.860) {$v_\parallel$: $N_v$ DG segments, GLL nodes; each (node, row) streams along its own line\\ at rate $v_\parallel g(r)$, Figures~\ref{fig:algorithm} and~\ref{fig:strip}};
\end{tikzpicture}
\caption{One time step in velocity space at one node: the $\vpar$ DG
segments with their GLL nodes and the $\mu$ rows, on GLL nodes in
$w = \sqrt{2B_0\mu}$ from $w = 0$.  The grid is cut above the top drawn
row: the runs carry eight intervals in $w$ and the figure two.  The kick (step
5 of Algorithm~\ref{alg:step}) is drawn on three rows in the form of
Figure~\ref{fig:strip}: red is a node, the dashed red box its segment,
black its foot, the shaded box that segment shifted by
$-(E_\parallel - \mu\,\partial_s B)\,\Delta t$, which differs from row
to row.}
\label{fig:velocity}
\end{figure}
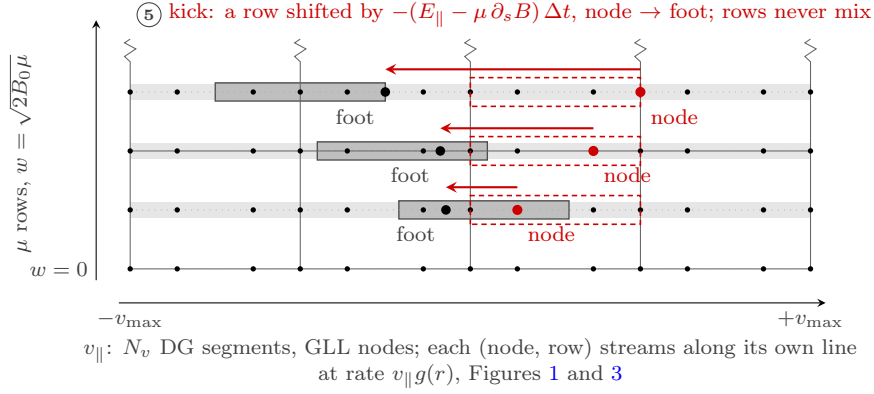

\begin{algorithm}[t!]
\caption{One time step of the composite, in the order the code applies
it; the numbers are those circled in Figures~\ref{fig:algorithm}
to~\ref{fig:velocity}, and the names are the figures'.  Every stage is an
independent per-$\mu$-row operation in velocity space.}
\label{alg:step}
\footnotesize
\begin{enumerate}
\item \textbf{Stream}, $\Delta t/2$.  For every pencil, a channel at one
  poloidal node, shift
  the DG-in-$\varphi$ representation along the field line by
  $\vpar\, g(r)\,\Delta t/2$, a static distance in wedges
  (Figures~\ref{fig:algorithm} and~\ref{fig:strip}).  Then the first
  half of the drift stage, $\Delta t/2$, as in step 7.
\item \textbf{Density.}  Take the velocity moment of $f$ at every
  station to form $\delta n$, then replace it by the midpoint prediction
  $\delta n - (\Delta t/2)\,\nabla\cdot\big((n_0 + \delta n)\,\mathbf{v}_E\big)$,
  with $\mathbf{v}_E$ from the previous step; on the concentric
  equilibrium $\nabla\cdot\mathbf{v}_E = B D\, v_{E,x}$.
\item \textbf{Field.}  Solve the polarization
  equation~(\ref{eq:field}) on each plane this rank owns.  The planes
  are independent, and the operator is assembled once.
\item \textbf{Filter} (optional).  Project $\phi$ onto the toroidal
  harmonics $|n| \le n_{\max}$ along the strip of
  Figure~\ref{fig:strip}, an all-gather of the planes over the
  ranks (Section~\ref{sec:cbc-nofastest}).
\item \textbf{Kick}, $\Delta t$.  Form $E_\parallel = -g(r)\,
  \partial_\varphi \phi$ along the line (the GLL derivative within the
  wedge) and remap every $\mu$ row along $\vpar$ by the affine map
  $v \mapsto \lambda v + \delta$ of the curvature term
  (Section~\ref{sec:rh}): the foot of node $v$ is $(v - \delta)/\lambda$
  with $\delta$ the shift $(E_\parallel - \mu\,\partial_s B)\,\Delta t$,
  the mirror force composed in, and $\lambda = \exp(-\kappa\,\Delta t)$,
  $\kappa = D\,\partial_y\phi$ evaluated at $\mathbf{x} +
  \tfrac12\Delta t\,\mathbf{v}_E(\mathbf{x})$, the midpoint of the
  trajectory the E$\times$B stage is about to transport.
\item \textbf{E$\times$B}, $\Delta t$.  Form
  $\mathbf{v}_E = \mathbf{b}\times\nabla\phi/B$ on every plane and move
  each (station, node) value to the foot of its characteristic,
  \begin{equation}
    \mathbf{x}_{\mathrm{foot}} = \mathbf{x} - \Delta t\,
      \mathbf{v}_E\!\left(\mathbf{x} - \tfrac{1}{2}\Delta t\,
      \mathbf{v}_E(\mathbf{x})\right),
    \label{eq:exbfoot}
  \end{equation}
  one midpoint iteration on the field frozen at the step midpoint
  (Figure~\ref{fig:plane}).  $\mathbf{v}_E$ does not depend on
  $(\vpar, \mu)$, so the foot and its interpolation weights are built once
  per (station, node) and read by every one of the $N_{ch}$ channels.
  A foot no triangle contains takes the equilibrium value, as the
  drift's does.
\item \textbf{Drift}, the second $\Delta t/2$.  Move each (node, channel)
  value along the drift's displacement, on the concentric equilibrium
  the node's vertical fiber (Figure~\ref{fig:algorithm}) by
  $(\vpar^2 + \mu B)\,D(x)\,\Delta t/2$ with the displacement factored
  into per-channel times per-node products, and evaluate the nodal basis at
  the foot in its containing triangle.  A foot outside the mesh takes
  the equilibrium value.
\item \textbf{Stream}, $\Delta t/2$, as in step 1.
\end{enumerate}
\end{algorithm}

The toroidal direction is the first MPI axis: each rank owns consecutive
wedges, and the stream's exchange is a nearest-neighbour shift along the
chain of ranks.  Everything else in the step is local to a plane, to a
(node, $\mu$ row) pair or to a fiber, apart from the one all-reduce of
surface sums that the flux-surface average adds.  The companion
paper~\cite{AdamsHPC2026} carries the $\mu$-row and plane axes of the
decomposition and their measured cost.

The filter of step 4 exists because the model has no gyroaverage, so
nothing cuts off high $k_\perp$ and the fastest-growing mode on a grid is
its highest, worst-resolved toroidal harmonic
(Section~\ref{sec:cbc-nofastest}).  The filter restricts the field, not
$f$, to $|n| \le n_{\max}$ before the kick and the E$\times$B stage use
it, so unresolved harmonics are never driven and no dissipation is added.
Its inner product weights each distinct station angle equally, because
the sampled harmonics are not orthogonal under the GLL quadrature.  In the
production configuration of Section~\ref{sec:cbc} (48 stations,
$n_{\max} = 5$) a kept harmonic is reproduced to $1.0\times10^{-15}$, and
no harmonic from $n = 6$ to the grid's 16 leaks more than
$4.2\times10^{-15}$.

The implementation is PETSc~\cite{PETSc2024} with
Kokkos~\cite{TrottKokkos2022}, as a library with one source file per
stage of the step.

\subsection{The equilibrium}
\label{sec:equilibrium}

The equilibrium is a canonical Maxwellian $f_0(p_\varphi, E, \mu)$, a
function of the invariants of the unperturbed
orbits~\cite{Angelino2006,Idomura2008}, as in ORB5~\cite{Lanti2020} and
GYSELA~\cite{Grandgirard2016}, with tanh-localized density and
temperature profiles in the style of global GENE.  The profiles'
normalized gradients, the \emph{label gradients} $R_0/L_n$ and $R_0/L_T$,
are exact at the reference surface $r_0$, and the profiles flatten beyond
${\sim}2\sigma$.  The profiles are evaluated at a canonical radius
$r_c = \psi^{-1}(\psi(r) + \Delta\psi_{\mathrm{kin}})$.  The label must be
an exact invariant of the flow the code integrates, so $p_\varphi$ is
derived from that flow, not taken from the textbook definition.  The
model flow of Section~\ref{sec:method}, rigid-$q$ streaming, the mirror
kick and the vertical drift, conserves
\begin{equation}
p_\varphi = \psi(r) + \frac{B_0 R_0}{B}\,\vpar ,
\qquad
\psi'(r) = B_0 R_0\, r\, \iota(r)\, g(r) ,
\label{eq:pphi}
\end{equation}
exactly and for every field model in the code, because the factor
$(\dd B/\dd x)/B^2$ common to the mirror force and the drift cancels
between them.

The cancellation extends to the perturbed flow.  The
E$\times$B displacement changes $(B_0R_0/B)\vpar$ at the rate
$\vpar B_0R_0 D\,\partial_y\phi/B$, and the curvature term
$\dot{\vpar} = -\vpar D\,\partial_y\phi$ changes it by the negative of
that, for every $B(x)$ model and with no expansion.  What is left is the
torque $-\partial\phi/\partial\zeta$, which is zero for a zonal field.
The flow does not conserve the textbook relation $\psi' = r B_\theta$,
because a rigid-$q$ line is not that equilibrium; the two differ by the
pitch factor $R_0 g(r) = 1/\sqrt{1 + (r\iota/R_0)^2}$, 0.8\% at $r_0$ and
0.7\% at the edge for the profile of Section~\ref{sec:cbc}.

A guiding-centre orbit integrator gates the invariant: the drift of
$p_\varphi$ along trapped and passing orbits falls by $2^5$ per halving of
the integrator's step.  With uniform $B$ or zero label gradients the
labeling reduces bitwise to the identity, so it leaves unchanged every
gate of Section~\ref{sec:ladder} that runs in either limit.  Without the
drift the invariants of the integrated flow are $(r, E, \mu)$, so a
drift-free run keeps $f_0(r, E)$, its exact equilibrium.

The labeling is what makes an unperturbed toroidal run converge.  Under
$f_0(r,E)$ the state deviation at $t = 10$ is the equilibrium relaxing,
and it does not move with resolution: $4.2\times10^{-2}$ at two and three
poloidal refinements alike.  Under the canonical $f_0$ it is
discretization error, $4.3\times10^{-3}$ at the production velocity grid
and three refinements, eight times below $f_0(r,E)$'s $3.5\times10^{-2}$
on the same grid.  The effective gradient is not the label gradient, and
at the normalized gyroradius of the demonstration the difference is order
one (Section~\ref{sec:cbc-dilution}).

\section{Geometry and the streaming operator}
\label{sec:geometry}

The mesh is an unstructured triangulation of the poloidal disk, extruded
in $N_\phi$ wedges whose GLL stations are copies of the plane
(Figure~\ref{fig:algorithm}).  The stream shifts each pencil along its
field line.  On circular concentric surfaces the pitch $\iota(r)$ is
constant per line, so a half-step transfer maps station $\varphi$ to
$\varphi - \Delta\varphi$, with $\Delta\varphi = \vpar\, g(r)\,\Delta t/2$,
and rotates the poloidal footprint by $\iota(r)\,\Delta\varphi$, where
$g(r) = \dd\varphi/\dd\ell$ is the rate along the line.  Because $g$ is
constant per line, the shift is a function of (channel, node) alone, and
every map in the step is built once except the E$\times$B stage's, which
is rebuilt every step (step 6).  The backtrace sits behind one interface
with two implementations: magnetic-coordinate, exact for the concentric
equilibrium, and Runge--Kutta, which integrates the field line of the
supplied $\psi(R,Z)$ on the shaped equilibrium (Section~\ref{sec:shaped}).
A new geometry changes the trace, not the transfer.

The stream is a map, not a stencil: the field line through a node lands
inside a triangle of the upstream plane, and the transfer evaluates the
nodal basis there, pointwise rather than as a projection.  This is the FCI
construction of Hariri and Ottaviani~\cite{Hariri2013} and the field-line
map of GRILLIX~\cite{Stegmeir2016,Stegmeir2018}, carried into a
semi-Lagrangian kinetic solver.  No parallel operator is discretized on a
misaligned grid, so no field-aligned coordinate becomes singular at the
magnetic axis or at an X-point, and no misaligned parallel operator leaks
truncation error into the perpendicular dynamics.  Field-aligned codes
can avoid that singularity in how they compute their
geometry~\cite{Shukla2026}.  Misalignment costs
instead the interpolation at the foot, a perpendicular numerical
diffusion that couples neighbouring field lines~\cite{Stegmeir2017}.

We measure that cost rather than bound it.  The in-plane interpolation
error converges at about order $p$ in the mesh spacing, one order below
the nodal interpolation's own, and the resolution budget of
Section~\ref{sec:cbc} accounts for that order.  The pointwise evaluation
does not conserve mass: the toroidal Landau gate drifts by
$1.4\times10^{-10}$ over two damping times
(Section~\ref{sec:ladder-landau}).  The dissipation-floor rung of
Section~\ref{sec:ladder-floor} runs a known damped rate through the full
chain on the working mesh and reads the excess damping at 1.6\% of the
growth rate the chain must resolve.

Three gates check the stream.  A global polynomial in space translates to
machine precision under a uniform displacement.  The $\varphi$-interpolation
order is measured under refinement.  The result is bitwise invariant
across toroidal process counts, a strict test of the exchange logic,
because the toroidal decomposition reorders no sum in the state's path.

\section{The verification ladder}
\label{sec:ladder}

Each operator entered the code behind a gate, an assertion with a
committed reference value that is checked by machine on every change
(\ref{sec:methodology}).  \emph{Degenerations} assert that in a parameter
limit the new operator reduces algebraically to an already-verified one,
bitwise where the arithmetic permits.  \emph{Physics gates} assert a
measured rate against a reference computed independently of the code.
Table~\ref{tab:ladder} lists the physics gates, and
Sections~\ref{sec:shaped} and~\ref{sec:ladder-cons} give the field path on
the shaped equilibrium and the conservation of the split step.  The two
tests with published references that exercise the whole step on the
zonal component, the geodesic acoustic mode and the Rosenbluth--Hinton
residual, are in Section~\ref{sec:rh}.

\subsection{The operator gates}
\label{sec:ladder-landau}
\label{sec:ladder-bounce}
\label{sec:field}
\label{sec:ladder-drift}
\label{sec:ladder-itg}
\label{sec:ladder-floor}

\begin{table}[ht!]
\centering
\caption{The physics gates of the operator ladder, in the order the
operators entered the code.  The Landau, slab-ITG and dissipation-floor
gates run with uniform $B$, the others on the circular concentric
equilibrium.  Every reference is computed by a tool distributed with the
code, against the code's own field model and, where it matters, its own
quadrature.}
\label{tab:ladder}
\footnotesize
\begin{tabular}{@{}>{\raggedright\arraybackslash}p{0.2\linewidth}>{\raggedright\arraybackslash}p{0.3\linewidth}>{\raggedright\arraybackslash}p{0.42\linewidth}@{}}
\toprule
gate & reference & measured \\
\midrule
toroidal parallel Landau damping & kinetic dispersion root~\cite{FriedConte1961} & rate within 1.44\% (1.25\% with the radius-dependent $k_\parallel$); mass drift $1.4\times10^{-10}$ over $t = 30$ \\
bounce and transit & elliptic-integral period~\cite{HelanderSigmar2002} in the code's $B$ & 0.9--3.6\% on both branches; splitting order 2.8 (Strang's 2) \\
mirror force & $\mu = 0$ damping rate on the axis surface & $1.4\times 10^{-13}$ relative \\
polarization solve & manufactured solution~\cite{Roache2002}, order $p+1$ & orders 2.99 ($P_2$) and 3.99 ($P_3$); $\rho_s^2 \to 0$ nodal limit $2.9\times10^{-13}$; GAMG and Cholesky agree to $1.5\times10^{-15}$ \\
drift orbits & guiding-centre ODE ensemble in the code's $B$ & banana and passing excursions within 1.8\% and 5.5\% (2.9\% at $P_3$) \\
slab ITG~\cite{Coppi1967} & matched discrete root of the radial eigenproblem & rate 2.68\% below the root \\
dissipation floor & a known damped rate, budget $\gamma_{\mathrm{target}}/10$ & excess decay $1.50\times10^{-3}$ against $9.4\times10^{-3}$ \\
\bottomrule
\end{tabular}
\end{table}

Two of the references need a sentence.  The slab-ITG reference is the
discrete root at the run's velocity resolution, not the continuum root:
at the operating point the resonance width is comparable to the $\vpar$
node spacing, and the velocity quadrature moves the growth rate by
$+11.4\%$, a gap that belongs to the discrete system.  The dissipation
floor also sets a resolution limit: a damped rate is observable only
until $\vpar$ filamentation reaches the grid scale at
$t \approx \pi/(k_\parallel h_v)$, the recurrence time of a discrete
velocity grid~\cite{ChengKnorr1976}, and doubling $N_v$ doubles that time.

\subsection{The field path on a shaped cross section}
\label{sec:shaped}

The polarization solve, the flux-surface average and the diagnostics read
the mesh through its nodes, cells and labels, with no assumption about the
shape of the flux surfaces.  On the up-down asymmetric Solov'ev
equilibrium of Figure~\ref{fig:algorithm} (inverse aspect ratio $0.36$,
elongation $1.7$ and $1.9$, triangularity $0.25$ and $0.45$, no X-point),
the manufactured solution of the polarization solve converges at the
circle's order on the flux-aligned ladder L1 to L5: 2.98 at $P_2$ against
the concentric control's 2.99 over the last level, and 3.82 against 4.00
at $P_3$, which enters its asymptotic range later on the elongated cells.
Conjugate gradients with algebraic multigrid (PETSc GAMG) take no more
iterations than on the control at the same node count (56 against 63 at
L5, $P_2$, to $10^{-12}$).  The flux-surface average's projector is
idempotent on the shaped labels to $2.2\times 10^{-16}$.

The parallel path reads the equilibrium the same way.  The field line
traced through the Runge--Kutta path of Section~\ref{sec:geometry}
leaves its flux surface only by the integrator's truncation, converging at
fourth order in the substep, and its toroidal transit reproduces an
independent integration of the same $\psi$ to nine digits.  The drift
stage carries the full guiding-centre displacement, whose component
across the vertical reaches 11\% on this equilibrium, and reproduces a
linear state at its own foot to $1.6\times 10^{-15}$.  The geodesic
acoustic frequency on this equilibrium is in Section~\ref{sec:gam}.

\subsection{Conservation under joint refinement}
\label{sec:ladder-cons}

The split step carries the invariant $W = K + U$, with
$K = \sum_q w_q f\,(\vpar^2/2 + \mu B)$ the full-$f$ kinetic energy in the
flux-tube measure of Section~\ref{sec:measure} and
$U = \tfrac{1}{2}\int n_0 \phi\, \delta n$ the polarization and
adiabatic-electron energy of the same solve.  We report the maximum of
$|W(t) - W(0)|/W(0)$ over the run, because the drift is not monotone.
Table~\ref{tab:cons} refines the toroidal and velocity axes together at
fixed physical time, with and without the zonal seed of
Section~\ref{sec:rh}; at $\alpha = 0$ the initial state is an exact
steady state of the continuum model, so every change in $W$ is
discretization error.

We found that the transport converges and its coupling to the field does
not.  Without the seed the energy drift falls $167\times$ over the two
joint refinements, from $1.80\times 10^{-3}$ to $1.08\times 10^{-5}$.  With
the seed it falls $2.1\times$ and then stalls at $5.2\times 10^{-4}$, which
neither halving the time step nor refining the poloidal mesh moves.  The
plateau is not the seam between the kinetic stages, which exchange energy
in the nodal quadrature, and the Galerkin field solve: removing the seam
leaves the seeded drift $128\times$ the $\alpha = 0$ floor.  Run to
$T = 100$, the seeded energy drift does not exceed its maximum by
$T = 50$, so it is an excursion that damps with the geodesic acoustic
oscillation driving it, not an accumulation.

The mass behaves the other way: it is flat under refinement
($8.6\times10^{-5}$ to $1.04\times10^{-4}$ across Table~\ref{tab:cons}) and
grows in time, from $9.1\times 10^{-5}$ by $T = 50$ to $2.6\times 10^{-4}$
by $T = 100$.  The transfer along the field line and the shift in $\vpar$
conserve mass under the $L^2$ projection; the in-plane stages evaluate at
their feet pointwise and change the mass by their sampling error.  A
flux-form variant of the drift stage, an experimental option, integrates
every cell's departure region exactly and carries the flux through the
open wall as a column of its own.  On the run above that wall flux is
$6\%$ of the pointwise drift's change, so the growth is in-plane sampling
error, not a flux through the wall.  With the drift on the variant the
maximum mass drift falls from $2.6\times10^{-4}$ to $1.5\times10^{-4}$,
and the energy drift rises from $5.2\times10^{-4}$ to $1.4\times10^{-3}$:
the pointwise drift's sampling error damps the perturbation that the
kinetic-field exchange otherwise accumulates.

\begin{table}[ht!]
\centering
\caption{Conservation under joint refinement of the toroidal and velocity
axes at fixed $T = 50\,a/c_s$.  Each entry is the maximum relative drift
over the run.  Seeded runs carry the zonal initial perturbation at
amplitude $\alpha = 1.6\times 10^{-2}$; at $\alpha = 0$ the initial
state is an exact steady state of the continuum model.}
\label{tab:cons}
\footnotesize
\begin{tabular}{llcccc}
\hline
 & & \multicolumn{2}{c}{$W$ drift} & \multicolumn{2}{c}{mass drift} \\
\cmidrule(lr){3-4}\cmidrule(lr){5-6}
wedges & $\vpar \times \mu$ & seeded & $\alpha = 0$ & seeded & $\alpha = 0$ \\
\hline
2 & $24\times12$ & $1.12\times10^{-3}$ & $1.80\times10^{-3}$ & $1.04\times10^{-4}$ & $2.23\times10^{-5}$ \\
4 & $48\times24$ & $5.45\times10^{-4}$ & $5.45\times10^{-5}$ & $8.57\times10^{-5}$ & $2.89\times10^{-6}$ \\
8 & $96\times48$ & $5.23\times10^{-4}$ & $1.08\times10^{-5}$ & $9.14\times10^{-5}$ & $2.37\times10^{-6}$ \\
8, $\Delta t/2$ & $96\times48$ & $6.11\times10^{-4}$ & --- & $1.20\times10^{-4}$ & --- \\
\hline
\end{tabular}
\end{table}

\section{Linear toroidal ITG on the canonical equilibrium}
\label{sec:cbc}

The configuration throughout this section is as follows.  The
equilibrium is the circular concentric one, $B = B_0R_0/R$ with
$B_0 = 50$ in units where $\rho_i = 1/B_0$, so
$\rho^* = \rho_i/a = 0.02$.  The safety factor $q$ is sheared
($q_0 = 0.854$, $q_a = 3.038$, $q(r_0) = 1.4$, $R_0/a = 2.78$).  The
label profiles are localized at $r_0/a = 0.5$ with $\sigma = 0.10$, and
$\tau = 1$.  The seeded mode is $(m,n) = (7,5)$, resonant at $r_0$, with
$k_\theta\rho_i = 0.28$.  The grid has four poloidal refinements ($P_2$,
spacing $h = 0.0625$ at $r_0$, $k_\theta h = 0.9$), $N_\phi = 16$,
$N_v = 12$ segments of degree 3 (48 nodes), $N_w = 24$ $\mu$ rows, and
$\Delta t = 0.1$.  The drift and E$\times$B are active, and there are
twenty radial bins.  The Cyclone Base Case label gradients are
$(R_0/L_n, R_0/L_T) = (2.2, 6.9)$.

\subsection{A diagnostic result on full-$f$ equilibria}
\label{sec:cbc-diagnostic}

A local Maxwellian $f_0(r, E)$ is constant along the parallel
characteristic but not along drift orbits, so with drifts and gradients
it is not a toroidal equilibrium~\cite{Angelino2006}; it relaxes, in
$(m,n) = (0,0)$.  At $B_0 = 50$ the banana width is two thirds of the
gradient-layer width, so the relaxation is order one on the profile.  At
labels $(2.2, 14)$ its field energy reaches $1.6\times10^{-3}$ by $t = 4$,
$4\times10^{5}$ times the $4.2\times10^{-9}$ of the seeded perturbation at
$\alpha = 10^{-4}$.  The growth rates of the radially binned RMS field at
$\alpha = 0$, $10^{-6}$ and $10^{-4}$ agree to $2\times10^{-4}$ over
$t \in [20, 60]$ in the four bins at the resonant surface, where the
projected mode below grows at 0.09 to 0.12.  An observable that sums over
helicities therefore reports what grows without the seed, not the seeded
mode.

The complex single-helicity projection isolates the seeded mode.  Per
radial bin, $a_b(t) = \langle E_\parallel\, e^{-i(n\varphi - m\theta)}
\rangle$ over the bin's nodes and all stations, with
$\gamma = \dd \ln|a_b|/\dd t$ and $\omega_r = -\dd \arg a_b/\dd t$
(phase unwrapped).  The projection is orthogonal to the $(0,0)$
background: on the discrete mesh a purely radial field of the mode's
amplitude projects to at most $10^{-15}$ of the projected mode, and a
synthetic single mode is recovered in amplitude and phase to
$2.5\times 10^{-13}$.

\begin{figure}[ht!]
\centering
\includegraphics[width=0.85\linewidth]{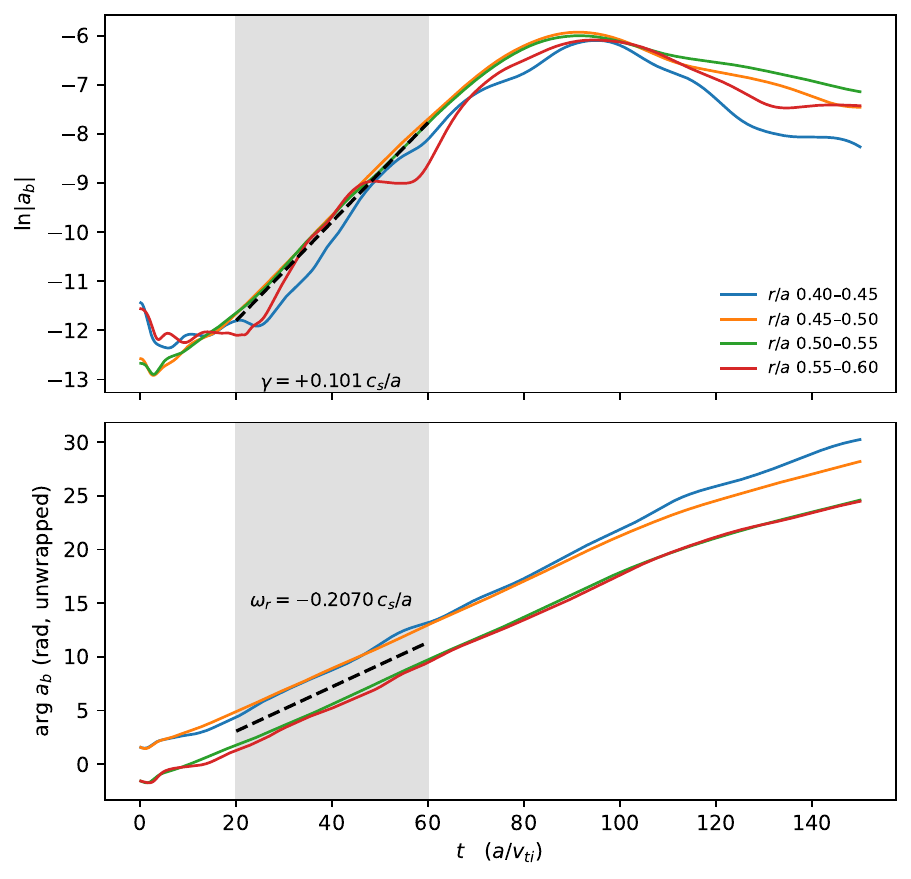}
\caption{Radial coherence of the projected $(m,n) = (7,5)$ mode at
labels $(2.2, 14)$ under $f_0(r,E)$ (the diagnostic result of
Section~\ref{sec:cbc-diagnostic}).  Log amplitude (top) and unwrapped
phase (bottom) of the
complex projection $a_b(t)$ for the four radial bins spanning
$r/a = 0.4$--$0.6$; the resonant surface $q(r_0) = 1.4$ sits at the
boundary between the middle two.  The shaded band is the fit window
$t \in [20, 60]$; the dashed guides carry the four-bin mean slopes.
Toroidal filter at $n_{\max} = 5$.  Per-bin fits:
$\gamma = \{0.111, 0.101, 0.098, 0.093\}$,
$\omega_r = \{-0.222, -0.201, -0.202, -0.204\}$ in $c_s/a$.  The
binned RMS field on the same run, which sums over helicities, decays at
0.011 to 0.019 over the same window (Section~\ref{sec:cbc-diagnostic}).}
\label{fig:coherence}
\end{figure}

At the CBC labels $(2.2, 6.9)$ the projected mode under $f_0(r,E)$ does
not grow, and its phase is not coherent across the bins at the resonant
surface.  At labels $(2.2, 14)$, with the toroidal filter at
$n_{\max} = 5$, it grows and is radially coherent
(Figure~\ref{fig:coherence}).  Over $t \in [20, 60]$, 3.8 e-foldings, the
four bins at the surface give rates of 0.093 to 0.111 and one frequency,
$-0.201$ to $-0.222$; without the filter the rates are 0.088 to 0.117.
At $t = 60$ the $n = 5$ harmonic of $\phi$ has the ballooning structure
of the toroidal branch, 2.5 times stronger on the outboard side than on
the inboard side.  The rate is above the canonical equilibrium's 0.065
at the same labels (Section~\ref{sec:cbc-demo}).  That is consistent with
$f_0(r,E)$ carrying the label gradients themselves until it relaxes,
where the canonical equilibrium carries the effective $(-0.79, 6.46)$ of
Section~\ref{sec:cbc-dilution}.

The rate, however, is not stationary.  Fitted to the four-bin mean over
successive 15-unit windows, it holds 0.087 to 0.114 through $t = 75$,
then falls to 0.052, 0.027 and $-0.018$, while the canonical run at the
same labels, with the same filter, holds 0.061 to 0.069 over the same
windows through $t = 105$.  The local rate starts to fall at an amplitude
1.6 times below the one at which the canonical mode starts to slow, which
suggests that the relaxation, by flattening the gradients, ends the
growth.  The projection, not the resolution, is what makes the linear
mode observable, and the canonical equilibrium is what makes its rate
stationary.

\subsection{Label gradients and effective gradients}
\label{sec:cbc-dilution}

The canonical labeling evaluates the profiles at the canonical radius,
so the flux-surface moments of $f_0$ are the label profiles convolved
with the distribution of label offsets, and the gradients that drive the
mode are the moments'.  We report both.  The effective gradients are
least-squares fits of $\ln n$ and $\ln T$ over $|r - r_0| \le \sigma/2$ to
the discrete moments of the initial condition, on the production node set
and velocity quadrature; they are properties of the initial condition,
not of the dynamics.  At the configuration above, the CBC labels
$(R_0/L_n, R_0/L_T) = (2.2, 6.9)$ give effective gradients
$(0.330, 3.276)$, and the labels $(2.2, 14)$ give $(-0.789, 6.460)$: the
density gradient inverts, and the temperature gradient is about half its
label.

The map from labels to effective gradients has a closed form.  To first
order in the labels, accurate to 1--5\% at every label used here, the
effective gradients are $\mathrm{sech}^2(\Delta/\sigma)$-weighted
Maxwellian moments of the label offset $\Delta$,
\begin{align}
R_0/L_{n,\mathrm{eff}} &= \phantom{-}\frac{R_0}{L_n}\,
  \big\langle \mathrm{sech}^2 \big\rangle
  - \frac{R_0}{L_T}\, \big\langle (E - \tfrac32)\,\mathrm{sech}^2 \big\rangle , \\
R_0/L_{T,\mathrm{eff}} &= -\frac{R_0}{L_n}\,\tfrac23
  \big\langle (E - \tfrac32)\,\mathrm{sech}^2 \big\rangle
  + \frac{R_0}{L_T}\,\tfrac23
  \big\langle (E - \tfrac32)^2\,\mathrm{sech}^2 \big\rangle ,
\end{align}
with $E$ the energy in units of $T(r_0)$ and the averages over the
Maxwellian at the node.  An implementation independent of the code, on
the code's node set and velocity quadrature, reproduces the code's
effective gradients to every printed digit on 22 configurations, and the
map inverts, giving the labels for any target effective gradient.

The map has two consequences.  First, the orbit average is not a small
correction at $\rho^* = 0.02$: the RMS label offset at $B_0 = 50$ is
0.100, equal to $\sigma$, and it falls as $1/B_0$.  Second, at
$\rho^* = 0.02$ the canonical-Maxwellian class contains no CBC point.  The
labels whose moment gradients read $(2.2, 6.9)$ are $(8.50, 19.20)$, and
that equilibrium carries a sheared parallel flow of Mach 0.3--0.5.  Its
energy distribution's second moment departs from a Maxwellian's by
$K = \langle E^2\rangle/\langle E\rangle^2/(5/3) - 1 = 0.22$ averaged over
the gradient layer and $0.48$ at its worst node ($K = 0$ for any
Maxwellian).  The CBC gradient is therefore quoted below as both the
label and the measured effective value, and the comparison with published
CBC rates is a $\rho^*$ ladder (Section~\ref{sec:cbc-ladder}) rather than
a point.

\subsection{The demonstration}
\label{sec:cbc-demo}

At labels $(2.2, 14)$, effective $(-0.79, 6.46)$, and perturbation
amplitude $\alpha = 10^{-4}$, we measure $\gamma = 0.065 \pm 0.001\ c_s/a$
and $\omega_r = -0.356\ c_s/a$, and the rate is stationary.  The fit
window is $t \in [25, 80]$, 3.6 e-foldings, ending before the mode of
Section~\ref{sec:cbc-nofastest} overtakes the run without the filter
near $t = 100$.  Table~\ref{tab:demo} gives the per-bin fits from the run
without the toroidal filter and from its repeat with the filter at
$n_{\max} = 5$; the two agree bin by bin to 0.5\%.  Two measurements
support the rate.  The six bins from $r/a = 0.33$ to $0.58$ span $0.0636$
to $0.0661$ without the filter, a half-range of 2\% (the $\pm0.001$
above), and sub-windows of 15 to 20 units within the fit window hold
between 0.058 and 0.068 in the bins at the surface.

\begin{table}[ht!]
\centering
\caption{Per-bin fits of the projected $(7,5)$ mode, labels $(2.2, 14)$,
window $t \in [25, 80]$, in $c_s/a$; $\omega_r$ in the convention of
Section~\ref{sec:cbc-diagnostic}.  Without and with the toroidal
filter ($n_{\max} = 5$) of Section~\ref{sec:splitting}.}
\label{tab:demo}
\small
\begin{tabular}{@{}lccccccc@{}}
\toprule
bin centre $r/a$ & 0.325 & 0.375 & 0.425 & 0.475 & 0.525 & 0.575 & 0.625 \\
\midrule
$\gamma$, no filter   & 0.0655 & 0.0661 & 0.0661 & 0.0649 & 0.0638 & 0.0636 & 0.0642 \\
$\gamma$, filter      & 0.0654 & 0.0661 & 0.0662 & 0.0652 & 0.0640 & 0.0638 & 0.0644 \\
$\omega_r$, no filter & $-0.355$ & $-0.357$ & $-0.356$ & $-0.356$ & $-0.357$ & $-0.357$ & $-0.356$ \\
\bottomrule
\end{tabular}
\end{table}

At the CBC labels $(2.2, 6.9)$, effective $(0.33, 3.28)$, the same
configuration does not grow: the projected amplitude decays by 3.5 to 3.9
e-foldings over 150 time units in the four bins at the surface.  The
effective drive there is $R_0/L_T = 3.3$, not 6.9: the map of
Section~\ref{sec:cbc-dilution}, seen directly in a run.

\subsection{Without a gyroaverage there is no fastest mode}
\label{sec:cbc-nofastest}

Without the filter the demonstration's $(7,5)$ rate stalls near
$t = 100$, and a field grows from $t \approx 80$ with the perturbation
amplitude set to zero.  What grows is the $n = 16$ ITG mode, the highest
toroidal harmonic the sixteen-wedge grid holds, seeded by discretization
error.  A field snapshot at $t = 110$ of the unperturbed run carries no
energy in $n = 1$ to $15$, and its poloidal number follows
$m = 16\,q(r)$ bin by bin, the resonant-harmonic structure of a
toroidal eigenmode at $k_\theta\rho_i = 0.9$.

The cause is the model, not the grid.  With $J_0 \to 1$ nothing cuts off
high $k_\perp$: the finite-Larmor-radius stabilization of the gyrokinetic
ITG mode~\cite{Romanelli1989} is absent, and the growth rate rises with
$n$ as far as the mesh can carry the mode.  Table~\ref{tab:gn} is the
measured ladder at the same drive.  The rate rises through $n = 8$.  At
$n = 16$ it is not converged: the two $N_\phi$ differ and the sub-windows
do not settle, at $k_\theta h = 2.75$ against a design target below 0.7
at $P_2$, which the $(7,5)$ mode, at 0.9, already exceeds.  The $n = 32$
mode is not representable at all, so its zero growth is not a cutoff.  At
this resolution we cannot determine whether the drift-kinetic model has a
maximum-growth $n$.  We can determine that the fastest mode on the
production grid is its worst-resolved one, so its rate is a
discretization number, and $N_\phi$ is a physics parameter rather than a
resolution parameter.

The filter of Section~\ref{sec:splitting} keeps only the harmonics the
mesh resolves.  Without it, the field of the seeded run at $t = 110$ is
62\% $n = 11$ and 24\% $n = 5$, the $n = 16$ mode beating with the seeded
one, and the $(7,5)$ rate over $t \in [100, 120]$ falls to between
$-0.03$ and $0.02$.  With it, the $(7,5)$ mode still grows over that
window, at 0.04 to 0.05, and over the fit window of Table~\ref{tab:demo}
the rate is unchanged to 0.5\%.

\begin{table}[ht!]
\centering
\caption{Measured growth rate against toroidal mode number at the
labels $(2.2, 14)$ of the demonstration, $\alpha = 10^{-4}$ seeded at
$(m,n)$, four poloidal refinements, $\Delta t = 0.1$; $k_\theta\rho_i$ is
taken at the mode's resonant surface and $k_\theta h$ is the poloidal
wavenumber times the mesh spacing at $r_0$.}
\label{tab:gn}
\begin{tabular}{@{}lccccl@{}}
\toprule
$(m,n)$ & $k_\theta\rho_i$ & $k_\theta h$ & $N_\phi$ & $\gamma$ ($c_s/a$) & \\
\midrule
$(7,5)$   & 0.28 & 0.9  & 16 & 0.065 & stationary \\
$(11,8)$  & 0.45 & 1.4  & 16 & 0.11  & sub-windows 0.08--0.17 \\
$(22,16)$ & 0.90 & 2.75 & 16, 32 & 0.12--0.17 & not converged \\
$(45,32)$ & 1.8  & 5.6  & 32 & $\approx 0$ & not representable \\
\bottomrule
\end{tabular}
\end{table}

\subsection{Comparison with the Cyclone Base Case}
\label{sec:cbc-ladder}

The comparison with published CBC rates~\cite{Dimits2000,Falchetto2008}
is a scan in $\rho^*$ at fixed $k_\theta\rho_i$ and fixed effective
gradients, whose $\rho^* \to 0$ limit is the local-Maxwellian CBC: $m$ and
$n$ scale with $B_0$ at $m/n = 1.4$, the poloidal refinement holds
$k_\theta h = 0.9$, and the labels come from the inversion of
Section~\ref{sec:cbc-dilution}.  We measured $B_0 = 100$
($\rho^* = 0.01$, mode $(14,10)$, labels $(3.97, 10.29)$, effective
$(2.24, 7.01)$) and $B_0 = 50$ (labels $(8.50, 19.20)$, effective
$(2.20, 6.90)$, the inversion confirmed to three digits on a production
run): $\gamma = 0.124 \pm 0.005$ and $0.136 \pm 0.003\ c_s/a$, with
$\omega_r = -0.226$ and $-0.222$.  The Cyclone benchmark's gyrokinetic
codes give $\gamma \approx 0.087$ and $\omega_r \approx -0.24\ c_s/a$ at
$k_\theta\rho_i = 0.28$, read from Fig.~1 of Dimits et
al.~\cite{Dimits2000} (0.11 and $-0.30$ in $v_{ti}/L_n$).  With
$J_0 \to 1$ our rates should sit above a gyrokinetic reference, and they
do, by $1.43\times$ and $1.56\times$; the GKNET/ORB5 benchmark of Okuda et
al.~\cite{Okuda2026} finds the CBC rates unchanged by the long-wavelength
field approximation, which places the expected departure in the
gyroaverage.  The next rung, $B_0 = 200$ with $N_\phi = 64$, is a
leadership-scale run outside this paper.

\section{Zonal flows}
\label{sec:rh}

An axisymmetric, radially varying potential first oscillates at the
geodesic acoustic frequency~\cite{Winsor1968}, the oscillation
Landau-damps, and a residual
survives~\cite{RosenbluthHinton1998,HintonRosenbluth1999}.  Such zonal
flows regulate ion-temperature-gradient turbulence~\cite{Lin1998}, and
their linear response is a standard cross-code test of gyrokinetic
codes~\cite{Biancalani2017}.  The oscillation and the residual are the
damped poles and the zero-frequency part of one linear response,
measured in the same runs, but they test different physics.  The
oscillation tests the geodesic coupling of the zonal flow to its
$m = \pm1$ sidebands and the passing-ion resonance
(Section~\ref{sec:gam}); the residual tests the neoclassical polarization
of the trapped and barely passing orbits (Section~\ref{sec:residual}).
Both exercise the drift stage, the mirror force, the parallel stream and
the energy exchange with $E_r$ together on the $(m,n) = (0,0)$ component
of the field solve, which no rung of Section~\ref{sec:ladder} does.

The residual is the ratio of the classical polarization to the classical
plus neoclassical polarization, $1/(1 + q^2\Theta/\epsilon^2)$, with
$\Theta \to 1.6\,\epsilon^{3/2}$ at large aspect ratio and small $k_r$
times the trapped-orbit width.  At the production point
($\rho_i = 0.02\,a$, $q = 1.4$, $\epsilon = 0.18$) the poloidal gyroradius
$\rho_p = q\rho_i/\epsilon$ is $0.16\,a$, so a zonal seed of wavenumber
$k_r = 12.6/a$ has $k_r\rho_p$ of order one, and the reference is the
finite-wavelength theory of Xiao and Catto~\cite{XiaoCatto2006}.  The
residual runs at $B_0 = 200$, where $k_r\rho_p = 0.49$.

\ref{app:residual} derives the neoclassical polarization of
Equation~(\ref{eq:dk}) as discretized and evaluates it by quadrature over
the code's own orbits: $\Theta = 0.1351$ at the test's point, against the
$0.1353$ of the Xiao--Catto expansion, and the model's single-surface
residual, with the seed's own phase mixing, is $0.1054$.  The measured
observable is a projection over the whole seed, whose potential is not a
scaled copy of its density, and \ref{app:residual-observable} carries the
same polarization through that projection: the reference for the
observable as measured is $0.085$.

\emph{Configuration and observable.}  The runs use the circular
concentric equilibrium and velocity grid of Section~\ref{sec:cbc} on the
concentric meshes L4 to L6 (L$n$ has $n$ poloidal refinements), flat
profiles and $\Delta t = 0.1\,a/c_s$ unless a table states otherwise.
The two-wedge zonal runs split the kick of Algorithm~\ref{alg:step} into
two half-steps about the E$\times$B stage with the nodal curvature
exponent, a second-order variant of the listed step; a control on the
listed step at L4 reproduces the geodesic frequency to $0.01\%$ and the
damping to $2\%$.

The initial condition multiplies $f_0$ by
$1 + \alpha\,e(r)$ with the envelope
$e = \cos k_r(r - r_0)\,\exp(-(r - r_0)^2/2\sigma_e^2)$, $r_0 = 0.5\,a$,
$\sigma_e = 0.15\,a$ and $\alpha = 1.6 \times 10^{-2}$.  The observable is
the flux-surface average of $\phi$ in 24 radial bins per step, projected
onto the seed's envelope with its disk mean removed and normalized to its
value after the first solve.  The radial field between neighbouring
surfaces, read through the same projection, agrees with it to within
12\% at every mesh and window.  A run at $\alpha = 0$ measures the
discretization's own zonal drift, 1\% to 8\% of the seeded signal over
the reading windows, which is subtracted; the subtracted residual is
flat to 4\% over a factor of 16 in the seed (\ref{app:gam}).

\emph{The flux-surface average.}  Without the average in
Equation~(\ref{eq:field}), the field equation for a zonal $\phi$ reduces
to $\phi = \tau\,\delta n$, and there is no residual to measure.  With
it, the per-plane operator of Section~\ref{sec:field} becomes
\begin{equation}
  B = A + M(I - P), \qquad P = E\,(E^\top M E)^{-1} E^\top M ,
  \label{eq:fsa}
\end{equation}
with $A$ the polarization stiffness, $M$ the $n_0$-weighted mass matrix
and $E$ the $N \times K$ operator that spreads $K$ flux-surface values
onto the nodes.  $P$ is the $L^2(n_0)$-orthogonal projection onto the
flux functions the mesh can represent; because the weight carries the
Jacobian, it is the torus's flux-surface average.  The average runs over
the poloidal surfaces \emph{and} the toroidal stations, so
$(m,n)=(0,0)$ is the one unshielded component and $P$ couples the
planes.  Because $P\mathbf{1} = \mathbf{1}$ and
$\mathbf{1}^\top M P = \mathbf{1}^\top M$, $B$ is symmetric positive
semi-definite and singular exactly on the constants.  The averaging
weight is the consistent mass matrix, because the lumped weight of a
$P_2$ vertex node is exactly zero.

The operator is time-independent, which makes the exact solve
affordable.  Writing $U = ME$, the Woodbury form needs $G = C^{-1}U$ and
the dense block $S = E^\top M E - U^\top G$, computed once in $K$ solves
of the operator $C = A + M$.  A step then adds one all-reduce of $K$
doubles for the surface sums, a dense $K \times K$ solve whose answer is
$\langle\phi\rangle$, and one correction field added to every station.
This is the two-field solve of XGC~\cite{Ku2009}, not its split solve.
Because $\rho_s^2 \propto 1/B(x)^2$ varies over the plane, the response to
a charge that is a flux function departs from a flux function by 0.25\%,
a sideband that a split solve re-averages away.

Four gates check the
implementation.  The projector annihilates $r^2\cos 2\theta$ on every
surface to $8\times10^{-17}$.  A manufactured solution whose flux-surface
average is $\cos(kr)$ converges through the full operator at orders 2.98
at $P_2$ and 3.92 at $P_3$.  With one surface the operator reproduces the
one without the average up to a constant ($4.4\times10^{-14}$), and with
one surface per node the pure polarization one ($6.7\times10^{-13}$).
And the Woodbury solution solves the assembled $B$ to
$8.3\times10^{-13}$ of $\|b\|$.

\emph{The energy term.}  A zonal potential has no parallel gradient, and
the drift displacement does not depend on $\phi$.  The coupling the
residual needs is therefore the energy exchange between the radial drift
and $E_r$, $-(f_M/T)\,\mathbf{v}_d \cdot \nabla\phi$ in the linearized
equation: the whole right-hand side for a zonal $\phi$, and the source of
both the neoclassical polarization and the geodesic acoustic oscillation.
Its grad-$B$ half is what the fixed-$\mu$ E$\times$B stage of
Section~\ref{sec:method} carries: by the chain rule,
$\mathbf{v}_E \cdot \nabla f_M|_\mu$ is the diamagnetic drive at fixed
energy plus that half.  Its curvature half is the last acceleration of
Eq.~\eqref{eq:dk}, $d\vpar/dt = -\vpar D\,\partial_y\phi$ on the
concentric equilibrium.  Over a step at frozen coefficients the parallel
equation is $\dot{v} = a - \kappa v$ with $\kappa = D\,\partial_y\phi$,
whose solution is the affine map $v \mapsto \lambda v + \delta$,
$\lambda = \exp(-\kappa\,\Delta t)$.

The kick therefore absorbs the term
exactly, with no new splitting error, by moving the foot of its own remap
from $v - \delta$ to $(v - \delta)/\lambda$.  The term matters away from
the zonal component too: in the toroidal ITG response it is the
$\omega_d$ in the numerator of the non-adiabatic part, so the rates of
Section~\ref{sec:cbc} carry it.

Two measurements test the term's discretization.  The projection
multiplies each fiber's $\vpar$ mass by $\lambda$, to $5\times10^{-14}$
of the sum before cancellation.  The compressibility of the E$\times$B
flow, read from the Jacobian of the stage's own foot map, and the
dilation together leave $7.3\times10^{-7}$, $1.1\times10^{-6}$ and
$6.8\times10^{-7}$ of the initial mass by $T = 50$ at $\Delta t = 0.1$,
$0.05$ and $0.025$, on the seeded run at the finest point of
Table~\ref{tab:cons}: under 1\% of that run's mass drift, with no trend
in the step.  The two half-kicks evaluate the exponent at both ends of
each trajectory, a trapezoidal rule for its Jacobian, so the remainder has
no first-order part; the whole kick applied before the transport with the
exponent at the node is first order in the same count.

\subsection{The geodesic acoustic mode}
\label{sec:gam}

The reference for this mode is the model's own root, which
\ref{app:gam} builds in two steps.  The first is the kinetic dispersion
relation of Zonca, Chen and Santoro~\cite{Zonca1996}, re-derived for
Equation~(\ref{eq:dk}) with the flux-surface-averaged electron response
in the $\epsilon \to 0$ limit, where the average is the cylinder's, and
with the curvature coefficient kept as a parameter $s$.  At the seed's
$q = 1.4$ it gives $1.018\,c_s/a$ at $s = 1$ and $0.720$ at $s = 0$.  At
a flat $q = 1.4$, with the electron closure in that same cylinder
average, we measured $1.009$ to $1.014$ and $0.725$, at two, four and
eight wedges and on two mesh families.

On the sheared profile $q = 0.854 + 2.184\,(r/a)^2$ the relation is local
in the zonal field, and its closed loop is a continuum: each radius rings
at its own root, and the zonal potential the run reports is the radial
integral of that winding field.  The reference for a sheared run is that
continuum, read through the same projection and fit as the run
(\ref{app:gam-continuum}).  In the cylinder's closure it reproduces the
run's dependence on $k_r$, the seed width and the shear to within 0.4\%
to 1.2\%, and the field between neighbouring flux surfaces rings at the
local root on every surface from $r/a = 0.53$ outward, to within 0.3\% to
3.6\% (\ref{app:gam-reading}).

The second step is the closure the code assembles.  The average in
Equation~(\ref{eq:fsa}) is the torus's, weighted by $R$, and the
$\epsilon \to 0$ relation does not carry it, so the model's root at the
seed's $\epsilon = 0.18$ is computed by a linear initial-value solver for
Equation~(\ref{eq:dk}) on one flux surface (\ref{app:gam-closure}).  In
the cylinder's average the solver reproduces the analytic relation, the
frequency to within 0.6\% and the damping to within 8.4\% for
$r/a \le 0.5$.  In the torus's average the frequency is $1.4\%$ lower at
the seed's centre and the damping $1.56$ times the cylinder's.  The
$R$-weighted average couples the $m = 0$ component to the $m = \pm 1$
sideband at order $\epsilon$ and holds a damping floor of $0.017$ to
$0.019\,c_s/a$ across the outer half of the seed, where the transit
resonance has fallen to $0.003$: the finite-aspect-ratio damping
enhancement of Gao~\cite{Gao2010}.

With run and reference in the same closure, the run tracks its model.
On the mesh one level coarser than the campaign's, at two wedges, the
projected signal fitted over $[0.05, 50]\,a/c_s$ gives $\omega = 0.9232$
and $\gamma = 0.0310\,c_s/a$ against the toroidal continuum's $0.9268$
and $0.0377$: the frequency is $0.4\%$ low and the damping $0.82$ of the
reference (Table~\ref{tab:gam-matched}).  Against the $\epsilon \to 0$
continuum the same run reads $0.969$ of it in frequency and $1.59$ times
it in damping, so that apparent over-damping is entirely the mismatch
between the run's closure and the reference's.

The damping is converged in the step: $\gamma = 0.03103$, $0.03100$ and
$0.03098\,c_s/a$ at $\Delta t = 0.1$, $0.05$ and $0.025\,a/c_s$.  The seed
amplitude moves it.  Over $\alpha = 10^{-4}$, $10^{-3}$ and
$4\times10^{-3}$ the damping agrees to $1\%$ at $0.90$ of the continuum,
and only the campaign's $1.6\times10^{-2}$ departs, by $9\%$
(Table~\ref{tab:gam-alpha}); we read that as the seed leaving the linear
regime for this observable.  At eight wedges the linear limit is $0.88$
to $0.94$ of the continuum by window.  The code therefore damps its own
geodesic-acoustic root at $0.88$ to $0.94$ of the continuum at every
toroidal resolution we ran.  We have not separated the remaining $6\%$ to
$12\%$ between the poloidal mesh and the continuum construction, which
couples no surface to its neighbours.

On the shaped equilibrium of Section~\ref{sec:shaped} we seeded the
surface with local inverse aspect ratio $0.16$, elongation $1.69$,
triangularity $0.13$ and $q = 1.60$, and measured $0.707$ and
$0.709\,c_s/a$ over $[0.05, 50]$ and $[0.05, 100]\,a/c_s$ at eight
wedges.  That is $3\%$ above the $0.688$ of the shaped dispersion
relation of Gao~\cite{Gao2010} at those parameters.  The relation
truncates at order $1/q^2$, omits the triangularity term and takes $R_0$
at the surface's centre rather than the axis, and those three omissions
place the expected run $3\%$ to $6\%$ above it.  This is the section's one
comparison to a published formula rather than to the model's own root,
because the initial-value solver is for circular surfaces.

\subsection{The Rosenbluth--Hinton residual}
\label{sec:residual}

The toroidal resolution moves the measurement most.
Table~\ref{tab:rh-tor} refines it alone, with the mesh, the velocity
grid, the step and the seed fixed and the reading window the last
quarter of a $T = 300\,a/c_s$ run.  From two wedges to eight the residual
rises 80\% at L4 and 61\% at L5, and it converges there: at L4 sixteen
wedges give $0.050$ and thirty-two $0.053$, and at L5 sixteen give
$0.070$ against eight's $0.069$, differences inside each reading's spread
over its window.  Doubling the parallel velocity resolution at sixteen
wedges moves the L4 reading by $8\%$, also inside the spreads.

At two wedges the poloidal sequence looks converged, $0.053$ at L5 and $0.047$
at L6 over $[100, 150]\,a/c_s$, but the L5 value lies $24\%$ below the
converged one, so two wedges do not resolve the measurement.  At two
wedges the time step moves the reading by 1\%, doubling the $\vpar$
resolution by 1.5\%, doubling the $\mu$ resolution by 0.3\%, and the
flux-surface basis at degree 1 in place of 2 by 2\%.

At a fixed eight wedges the poloidal mesh moves the measurement toward
the model's $0.085$: $0.061$ at L4, $0.069$ at L5 and $0.079$ at L6.  The
L6 value is read on an earlier window that reads about $6\%$ high, so on
a common window the three lie between $0.061$ and $0.074$.  The sequence
is monotone in the mesh with no sign of having stopped.  Its weaker end
is L4, where sixteen and thirty-two wedges give $0.050$ and $0.053$;
taking the L4 point at that converged value steepens the rise rather
than removing it.

An Eulerian discretization of the same model, built in the same code,
gives an independent reading.  It is discontinuous Galerkin, as Gkeyll
is in phase space~\cite{Mandell2020}, in $(R, Z, \vpar)$ on each $\mu$
row and axisymmetric, so it has no
toroidal resolution to converge, and it runs on the same meshes, velocity
grid, seed and step.  Over $[225, 300]\,a/c_s$ it reads $0.060$, $0.071$
and $0.070$ at L4, L5 and L6: it stops moving with the poloidal mesh at
$0.82$ of the model, where the semi-Lagrangian sequence, read on a common
window, still rises from $0.069$ to $0.074$.  Its residual is not
converged in the parallel velocity.  At a nearby configuration
($B_0 = 500$, $k_r\rho_p = 0.2$, the same seed shape) doubling the $\vpar$
nodes from 48 to 96 raises it by $16\%$, while doubling the $\mu$ rows
moves it by $1\%$; the semi-Lagrangian reading moves by $8\%$ with the
same doubling at L4 (Table~\ref{tab:rh-tor}).  On a common window the
two discretizations agree with the eight-wedge rows to within $6\%$ on
every mesh, and both lie below the model.  The parallel velocity
resolution is one candidate for that gap; the reference itself is
another (below).

\begin{table}[htbp]
\centering
\caption{The zonal residual against toroidal resolution, at fixed mesh,
velocity grid, step and seed, $k_r\rho_p = 0.49$, baseline-subtracted.
$N_\phi$ is the number of toroidal wedges, three GLL stations each.  The
residual is the projected observable over the last quarter of a
$T = 300\,a/c_s$ run, with the spread of the reading window; the lobe is
the same projection restricted to the six bins the seed's inner lobe
occupies, $r/a = 0.156$ to $0.355$, as a cumulative mean over $[30, T]$.
$^\ast$ is sixteen wedges with the parallel velocity resolution doubled,
to $96 \times 24$.  $^\dagger$ is read on the last quarter of
$T = 200\,a/c_s$ at $\Delta t = 0.05\,a/c_s$; the two L5 rows read on
that window give $0.074$, so the row is about $6\%$ high on this table's
window.  The model's prediction for the projected observable is $0.085$
(\ref{app:residual-observable}).}
\label{tab:rh-tor}
\begin{tabular}{llccccc}
\toprule
 & & & \multicolumn{4}{c}{lobe, $[30, T]$ for $T =$} \\
\cmidrule(l){4-7}
mesh & $N_\phi$ & residual & 50 & 70 & 90 & 110 \\
\midrule
L4 & 2  & $0.034 \pm 0.003$ & 0.078 & 0.057 & 0.034 & 0.004 \\
L4 & 8  & $0.061 \pm 0.007$ & 0.144 & 0.128 & 0.126 & 0.134 \\
L4 & 16 & $0.050 \pm 0.005$ & 0.161 & 0.145 & 0.139 & 0.144 \\
L4 & 32 & $0.053 \pm 0.009$ & 0.147 & 0.134 & 0.132 & 0.141 \\
L4 & 16$^\ast$ & $0.054 \pm 0.004$ & 0.151 & 0.141 & 0.139 & 0.127 \\
L5 & 2  & $0.043 \pm 0.004$ & 0.057 & 0.035 & 0.016 & $-0.005$ \\
L5 & 8  & $0.069 \pm 0.004$ & 0.127 & 0.121 & 0.129 & 0.120 \\
L5 & 16 & $0.070 \pm 0.005$ & 0.128 & 0.122 & 0.130 & 0.121 \\
L6 & 8$^\dagger$ & $0.079 \pm 0.011$ & 0.135 & 0.130 & 0.137 & 0.128 \\
\midrule
\multicolumn{2}{l}{the model} & 0.085 & 0.16 & 0.16 & 0.16 & 0.16 \\
\bottomrule
\end{tabular}
\end{table}

The seed's inner lobe separates the toroidal artifact from what remains.
The model shields that lobe least, to $0.16$ of its initial density
(\ref{app:residual-observable}).  At two wedges the run shields it away:
its cumulative mean falls from $0.078$ to $0.004$ of its seed at L4 and
to $-0.005$ at L5 (Table~\ref{tab:rh-tor}).  At eight wedges and above it
holds between $0.12$ and $0.16$ on two meshes, at three toroidal
resolutions and with the parallel velocity resolution doubled, so the
decay is an artifact of the toroidal resolution.  Nor does the model
itself approach slowly on these small-$\epsilon$ surfaces.  A radially
local initial-value solution on one flux surface (the solver of
\ref{app:gam-closure}) settles at $0.148$, $0.144$ and $0.135$ of its
seed on $r/a = 0.23$, $0.29$ and $0.35$, the Xiao--Catto level, by
$t = 60$ to $100\,a/c_s$, and holds it to $0.002$ through $t = 150$.

The shortfall in the residual itself does not go with the lobe.  The
runs that resolve the toroidal direction read $0.050$ to $0.070$ against
the model's $0.085$; on the eight-wedge comparison the prediction is
$1.39$ times the measurement on the coarsest mesh and $1.15$ times on the
finest.  On the semi-Lagrangian runs the remainder survives, within the
reading spreads, refinement of the time step, both velocity axes, the
toroidal direction, the flux-surface basis and the seed amplitude, but the
Eulerian reading above shows that the $\vpar$ refinement it survives is
not yet in the asymptotic range.  The field solve is excluded: the one-dimensional
polarization solve of \ref{app:residual-observable} reproduces the seed's
projected potential through the flux-surface projector to 0.3\%.

The reference is not an expansion: \ref{app:residual} evaluates the
model's polarization by quadrature over its own orbits, so the
truncation of the Xiao--Catto series does not enter.  What the reference
keeps is radial locality.  Equation~(\ref{eq:res-observable}) applies each
surface's enhancement to the potential on that surface, which neglects
the coupling between surfaces through the orbits, whose width is $0.17$
to $0.25$ of the radius on the inner surfaces, and the orbits that reach
the axis.  Global gyrokinetic codes bound where this approximation
holds.  For a long-wavelength seed, the residual of NEMORB meets the
Xiao--Catto long-wavelength limit below $\rho^* = 0.01$, but its comparison
at finite $k_r\rho_p$ stays open because the radial wavenumber of the
zonal potential changes during the run~\cite{Biancalani2014}.  The
residual of GYSELA approaches the Rosenbluth--Hinton formula as $\rho^*$
decreases, with the zonal potential's radial profile changing in time
against the theory's assumption~\cite{Grandgirard2016}.  In
quasi-symmetric stellarators the local and full-volume residuals agree at
moderate radial wavenumbers and differ at system-size
scales~\cite{Smoniewski2021}.  The runs here are at $\rho^* = 0.005$,
inside NEMORB's long-wavelength agreement, but at $k_r\rho_p = 0.49$, and
the seed's potential carries a whole-disk part at the system size
(\ref{app:residual-observable}), so the reference's locality remains a
candidate.  A scan
toward the local limit at fixed $k_r\rho_p$, raising $B_0$ and $k_r$
together, should move the locality's share of the gap and not the
velocity resolution's.

A third candidate is numerical, and it is testable.  With the drift on
the flux-form variant of Section~\ref{sec:ladder-cons}, and everything
else as in the L4, eight-wedge row, the lobe reads $0.160$ to $0.177$,
the model's $0.16$, and the plateau a window mean of $0.09$ to $0.12$,
where the pointwise control reproduces $0.061 \pm 0.007$.  The pointwise
drift's sampling error is a numerical dissipation that damps the
geodesic acoustic mode and lowers the residual.  The exact drift does not
settle the question.  On its trajectory the oscillation holds at $0.11$
of the seed through $t = 300$, where collisionless damping would have
removed it, and the energy defect is secular ($+1.8\times10^{-3}$ of
$W(0)$ by $t = 300$): the seam of Section~\ref{sec:ladder-cons} acts as
an energy source that the pointwise drift's dissipation had balanced.

\section{Performance}
\label{sec:performance}

The solver is fully device resident: on NVIDIA A100 nodes its state
stays in device memory across the step.  The companion
paper~\cite{AdamsHPC2026} covers the domain decomposition over the
toroidal angle, the poloidal plane and the $\mu$ rows.  Two numbers
locate the cost.  The linear toroidal ITG demonstration of
Section~\ref{sec:cbc-demo} runs 1,200 steps, to $t = 120\,a/c_s$, on one
A100 in under seven minutes, $332$\,ms per step.  A weak series on the
shaped equilibrium that refines all five phase-space dimensions
together, growing the work $16\times$ per point, holds a weak-scaling
efficiency (the one-GPU step over the step) of $57\%$ on 256 GPUs.

\section{Conclusion}
\label{sec:conclusion}

This paper presents a continuum drift-kinetic solver on unstructured
poloidal meshes, semi-Lagrangian in every transported direction and
discontinuous Galerkin along the field line and in $\vpar$, and the
verification ladder behind which every operator of its split step
entered.  Every rate is read against a reference computed independently
of the code.  The slab ITG rate is within $3\%$ of the matched discrete
root of its eigenproblem (Section~\ref{sec:ladder-itg}).  The toroidal
ITG rate on the canonical equilibrium is stationary, and at the
CBC-effective drive it falls $9\%$ between $\rho^* = 0.02$ and $0.01$
(Section~\ref{sec:cbc}).  The geodesic acoustic frequency is within
$0.6\%$ of the model's own continuum and its damping within $12\%$ in the
linear limit, with run and reference in the same electron closure
(Section~\ref{sec:gam}).  The Rosenbluth--Hinton residual rises with poloidal
refinement toward the model's own prediction.  An Eulerian discretization
of the same model on the same meshes stops moving with the mesh at $0.82$
of it and rises $16\%$ with doubled parallel velocity resolution.  The
candidates for the remaining gap are that resolution, the radial
locality of the reference, from which global gyrokinetic codes also
depart at finite $\rho^*$, and the drift stage's own sampling dissipation
(Section~\ref{sec:residual}).

The two invariants separate (Section~\ref{sec:ladder-cons}).  The energy
drift is an excursion that damps with the mode driving it and reaches no
larger value when the run doubles.  The mass drift grows with the run,
and no refinement of the toroidal, velocity or poloidal axis reduces it;
its growth is the drift stage's pointwise sampling error, whose wall flux
a flux-form variant of the stage puts at $6\%$ of its change.

The most urgent addition is collisions, which a nonlinear campaign needs
before anything else, because a damped mode is unobservable once velocity
filamentation reaches the grid (Section~\ref{sec:ladder-floor}).  This
discretization can take them on its own velocity grid, as a
discontinuous-Galerkin Landau operator with conservation by the weak
form, the route by which Gkeyll's DG Dougherty operator conserves density,
momentum and energy exactly~\cite{Hakim2020,Francisquez2020NF}.  Kinetic electrons are the largest addition and the one the edge
requires; the gyroaverage and the X-point mesh follow.  A conservative
in-plane remap ranks with them, because the mass defect is the one
defect the ladder measures that both grows with the run and survives
refinement of every axis.  The drift variant of
Section~\ref{sec:ladder-cons} is its first piece; it leaves each stage's
change of the mass as a computed flux through the wall without making the
defect small.  The parallel design and its scaling on GPUs are the
subject of the companion paper~\cite{AdamsHPC2026}.

\section*{AI-assisted implementation}

The operators of Section~\ref{sec:method}, the gate ladder of
Section~\ref{sec:ladder} and its reference tools were implemented, and
the text of every section drafted, with Anthropic's
Claude~\cite{claude} (Fable~5 and Opus~5 models), an AI code
assistant working under the author's direction; every section was edited
by the author.  What made the delegation safe was the gate of
\ref{sec:methodology} and not trust in the assistant: with a
bit-identical or tolerance-bounded acceptance criterion enforced on every
commit, a change that moves a gated result cannot be committed, and the
results trace to the committed artifact (\ref{app:artifact}).

\section*{Acknowledgments}

This work was supported in part by the U.S. Department of
Energy, Office of Science, Office of Advanced Scientific Computing
Research's Applied Mathematics Competitive Portfolios program under
Contract No. DE-AC02-05CH11231.  This
research used resources of the National Energy Research Scientific
Computing Center, a DOE Office of Science User Facility supported by the
Office of Science of the U.S. Department of Energy under Contract No.
DE-AC02-05CH11231 using NERSC award ASCR-ERCAP0039362.

\appendix
\section{Verification methodology}
\label{sec:methodology}

This appendix states the practices behind Section~\ref{sec:ladder}.

\emph{Committed reference values.}  Every gate asserts against a number
committed to the repository, and no change is committed unless the ladder
reproduces its committed values, bit-identically where the gate is a
degeneration and within stated tolerance where it is a physics rate.  A
commit that moves code and changes no arithmetic is also held to a byte
comparison of the physics output against the previous binary's, on a
committed set of configurations (51 at the time of writing).  A byte
comparison covers only the paths its configurations take, so every
option that is off by default carries a configuration that turns it on,
including options that are bit-identical by construction and therefore
invisible to a comparison that leaves them off.

\emph{Sharp degenerations.}  This is the primary tool.  A new operator
enters behind a parameter limit in which it must reduce
\emph{algebraically} to an already-verified one.  Uniform $B$ makes the
mirror force and drift vanish and the relabeling an exact identity.
Collapsing the perpendicular grid to one $\mu$ row at $\mu = 0$ makes the
two-velocity solver reproduce the one-velocity solver byte for byte.
Zero profile gradients reproduce the ungraded runs bitwise.  The new
machinery is added around the old path, not through it, so the
degeneration stays exact rather than holding to a tolerance.

\emph{Where the assertion is placed matters.}  The polarization solve's
$\rho_s^2 \to 0$ gate asserts the algebraic adiabatic response
\emph{nodally}, not on a downstream growth rate.  Because both sides of
the degenerate system are formed with the same weighted mass operator,
the identity holds for any quadrature and any $n_0$ profile.  Asserting
it at the nodes detected an error that a downstream check would have
averaged away.

\emph{Cross-build and cross-rank reproduction.}  The full ladder is
reproduced digit-for-digit between the CPU development build and the CUDA
production build, and bitwise across toroidal process counts.  Two practical
notes follow.  A host-side reference sum does not reproduce a device sum
that contracts $a b + c$ into a fused multiply-add, so a statement that
two sums agree identically must be a statement about two device
computations.  And a CPU run cannot detect host/device synchronization
errors, because host and device views alias on a host backend, which is
why the GPU pass is a separate gate rather than a convenience.

\emph{References are computed, not recalled.}  Every physics gate's
reference is generated by an independent tool run against the code's own
field model and, where it matters, the code's own quadrature: the
elliptic-integral periods, the guiding-centre ODE ensemble, and the
radial eigenproblem at matched velocity resolution.  The tools are
themselves gated.  The eigenproblem tool, for example, asserts its own
degenerations before it is used as a reference for the code.

\providecommand{\paperEq}[2]{Equation~\eqref{#1}}
\providecommand{\paperSec}[2]{Section~\ref{#1}}
\providecommand{\paperAlg}[2]{Algorithm~\ref{#1}}

\section{The geodesic-acoustic root of the model}
\label{app:gam}

The zonal test of \paperSec{sec:rh}{the Rosenbluth--Hinton section of the
paper} measures a geodesic acoustic oscillation before it measures a
residual.  The oscillation is the sharper comparison, a root of a published
kinetic dispersion relation rather than a limit formula.  This appendix
derives that relation for
\paperEq{eq:dk}{the drift-kinetic equation of the paper} with the
flux-surface-averaged electron response of
\paperEq{eq:fsa}{the flux-surface-averaged field operator of the paper},
evaluates it, and reads the runs against it.  The curvature coefficient is
carried as a parameter $s$, because the code runs with the curvature term on
($s = 1$) and off ($s = 0$) and both runs are compared here.

\subsection{The linearized model}
\label{app:gam-model}

Units are those of the code: $v_t^2 = T_i/m = 1$, $\phi$ in $T_i/e$,
$\tau = T_e/T_i$, $\Omega = B$ and $R = R_0$.  Linearizing
\paperEq{eq:dk}{the drift-kinetic equation of the paper} about
$f_0 = f_M(E)$ at flat profiles gives
\begin{equation}
\frac{\partial\,\delta f}{\partial t}
 + \vpar\nabla_\parallel \delta f
 + \mathbf{v}_d\cdot\nabla\,\delta f
 = -\frac{f_0}{T}\left[\, \vpar\nabla_\parallel\phi
   + \left(s\,\vpar^2 + \mu B\right) D\,\partial_y\phi \,\right] ,
\label{eq:gam-lin}
\end{equation}
with $D = -(\dd B/\dd x)/B^2$ as in the drift stage.  The right-hand side is
the energy exchange between the radial drift and $E_r$, and each half of it
is a stage of the split step.  Its $\mu B$ half is the fixed-$\mu$
E$\times$B advection of $f_0(\vpar, \mu B(\mathbf{x}))$, which the
semi-Lagrangian E$\times$B stage carries whenever it runs.  Its $\vpar^2$
half is the curvature term, $\dot{\vpar} = -\vpar D\,\partial_y\phi$ acting
on $\partial f_0/\partial\vpar$, which the affine foot of the acceleration
sweep carries and which $s$ multiplies.  The left-hand side does not depend
on $s$: the drift stage displaces every velocity channel by the full
$(\vpar^2 + \mu B)D\,\Delta t$ whatever the energy exchange does.

Write $\omega_d^{(s)} := k_r\sin\theta\,(s\vpar^2 + \mu B)/(\Omega R)$,
$\omega_d := \omega_d^{(1)}$, and split the response as
$\delta f = -f_0\phi + h$.  Then \eqref{eq:gam-lin} becomes
\begin{equation}
\left(\partial_t + \vpar\nabla_\parallel + i\,\omega_d\right) h
 = f_0\,\partial_t\phi
 + i\left(\omega_d - \omega_d^{(s)}\right) f_0\,\phi .
\label{eq:gam-h}
\end{equation}
At $s = 1$ the last term vanishes and \eqref{eq:gam-h} is the standard
drift-kinetic equation of every geodesic-acoustic derivation.

The field equation is quasineutrality with the polarization density
$n_0\rho_i^2\nabla^2\phi$, $\rho_i = v_t/\Omega$ and $\rho_s^2 = \tau\rho_i^2$,
and the electron response
$n_0(\phi - \langle\phi\rangle)/\tau$.  Its $m = 0$ component is
$k_r^2\rho_s^2\,\phi_0 = \tau\,\delta n_0/n_0$ and its $m = \pm 1$ component
is $(1 + \tau + k_r^2\rho_s^2)\,\phi_1 = (\tau/n_0)\int h_1$.  This is the
adiabatic-electron response of the textbook derivation, which takes the
$m \neq 0$ electron density as adiabatic and sets its $m = 0$ component to
zero~\cite{Biancalani2017}; the flux-surface average produces it rather than
departing from it.  The electrons shield the $m = 1$ sideband, which carries
the parallel gradient $k_\parallel = 1/(qR_0)$, and do not shield the $m = 0$
component, on which no parallel electron motion connects flux surfaces.  The
$(m,n) = (0,0)$ component that \paperEq{eq:fsa}{the flux-surface-averaged
field operator of the paper} leaves unshielded is exactly the one the
textbook derivation leaves unshielded, and the adiabatic-electron model of
the codes in the cross-code benchmark, $n_e = n_0 + n_0(\phi - \bar\phi)/\tau$
with $\bar\phi$ the flux-surface average, is the same
closure~\cite{Biancalani2017}.

\subsection{The dispersion relation}
\label{app:gam-disp}

The ordering is $k_r\rho_i = 0.063 =: \delta$ small, $\omega \sim v_t/R$,
and $\epsilon \to 0$ in the geometry, with the transit frequency
$\omega_t := \vpar/(qR)$ kept exactly rather than expanded in $1/q$;
finite $\epsilon$ enters at the end.  Expanding $h$ and $\phi$ in $\delta$
with $\exp(-i\omega t)$, the order-zero response is $h_0 = f_0\phi_0$, so
the zonal density is second order.  The order-one sideband is driven by
$-i\,\omega_d^{(s)} f_0\phi_0$, so the zonal-to-sideband leg carries
$a_s := s\,\vpar^2 + \vperp^2/2$, the combination the energy exchange
supplies: with $\phi_1 = \Phi\sin\theta$ and
$h_1 = h_1^+e^{i\theta} + h_1^-e^{-i\theta}$,
\begin{equation}
h_1^{\pm} = \left[\,\mp\frac{i\,k_r a_s\phi_0}{2\Omega R}
            + \omega\,\phi_1^{\pm}\right]\frac{f_0}{\omega \mp \omega_t} ,
\label{eq:gam-h1}
\end{equation}
and the $m = \pm 1$ field equation, whose $k_r^2\rho_s^2$ is order
$\delta^2$ and dropped, gives the sideband potential
\begin{equation}
\Phi = \frac{\tau\,k_r\,\phi_0\,M[a_s]}{(\Omega R)(1 - \tau N)} ,
\qquad
M[g] := \left\langle \frac{g}{\omega - \omega_t}\right\rangle ,
\qquad
N := \omega M[1] - 1 ,
\label{eq:gam-Phi}
\end{equation}
with $\langle\cdot\rangle$ the Maxwellian average.  The order-two zonal
density is the $m = 0$ part of
$-i\omega_d h_1 + i(1-s)\,\omega_{dc} f_0\phi_1$, with $\omega_{dc}$ the
curvature part of $\omega_d$: the drift advection of the sideband, so the
return leg carries the full drift $a_1$ at any $s$, and the last term of
\eqref{eq:gam-h} with $\phi_1$ in place of $\phi$, which supplies the
$(1-s)\langle\vpar^2\rangle$ below.  With
$K[g] := \omega M[g] = \langle g\,\omega^2/(\omega^2 - \omega_t^2)\rangle$
for even $g$, zonal quasineutrality closes the loop:
\begin{equation}
\frac{2\,\omega^2 R^2}{v_t^2}
 = K[a_1 a_s]
 + \frac{\tau\,K[a_s]\left(K[a_1] - (1-s)\langle\vpar^2\rangle\right)}
        {1 - \tau N} .
\label{eq:gam-disp}
\end{equation}
Every velocity integral in \eqref{eq:gam-disp} reduces to
$G_n(\zeta) = \langle \vpar^n/(\zeta - \vpar)\rangle$ with
$\zeta = qR\omega/v_t$, that is, to the plasma dispersion function
$Z(\zeta/\sqrt{2})$ with the Landau prescription.  The
first term is linear in $s$, because one of its legs carries $s$ and the
other does not; the second is quadratic in $s$.

In the large-$q$ limit $K \to \langle\cdot\rangle$ and $N \to 0$, and
\eqref{eq:gam-disp} becomes
\begin{equation}
\frac{2\,\omega^2 R^2}{v_t^2} = (4s + 3) + \tau\,(1 + s)^2 ,
\label{eq:gam-largeq}
\end{equation}
which at $s = 1$ is $\omega^2 = 2(7/4 + \tau)\,v_t^2/R^2$.  Note that the
textbook $\sqrt{7/4 + \tau}\;v_{ti}/R_0$ is written for
$v_{ti} = \sqrt{2T_i/m}$, while the code's Maxwellian is
$\exp(-v^2/2)/\sqrt{2\pi}$ and its $v_t$ is $\sqrt{T_i/m}$; in the code's
$v_t$ the same formula reads $\sqrt{2(7/4+\tau)}\,v_t/R_0$, a factor
$\sqrt{2}$ larger.  Expanding \eqref{eq:gam-disp} one order further in
$1/q$ at $s = 1$ gives
$\omega^2 = \omega_0^2\,[\,1 + (23/8 + 2\tau + \tau^2/2)/(q^2(7/4+\tau)^2)\,]$,
which is the explicit formula of Sugama and
Watanabe~\cite{SugamaWatanabe2006}.  In the notation of Zonca, Chen and
Santoro~\cite{Zonca1996} as printed in the cross-code benchmark of Biancalani
et al.~\cite{Biancalani2017}, write
$F_Z(x) = x(x^2+3/2) + (x^4+x^2+1/2)Z(x)$,
$N_Z(x) = x + (1/2+x^2)Z(x)$ and $D_Z(x) = (1/x)(1 + 1/\tau) + Z(x)$ at
$x = \zeta/\sqrt{2} = qR\omega/(\sqrt{2}v_t)$.  Then \eqref{eq:gam-disp} at
$s = 1$ is $x/q^2 + F_Z(x) - N_Z^2(x)/D_Z(x) = 0$, the dispersion relation of
that reference; we checked the identification term by term,
$K[a^2] = -4xF_Z$, $K[a] = -2xN_Z$ and $1 - \tau N = \tau x D_Z$.

The zonal test runs at $q(r_0) = 1.4$, where the $1/q$ expansion is 3.4\%
below the exact root, so we solve \eqref{eq:gam-disp} with the plasma
dispersion function instead.  The streaming rate carries the line-length
pitch factor, so the transit frequency is set by
$q_{\mathrm{eff}} = q\sqrt{1 + (r/qR_0)^2} = 1.4115$ at the seed.  The
exact roots there, at $\tau = 1$ and in units of $c_s/a$
($v_t/R_0 = 0.35997$), are $\omega = 1.018$ with $\gamma = 0.0139$ at
$s = 1$, and $\omega = 0.720$ with $\gamma = 0.0640$ at $s = 0$; the
large-$q$ limit \eqref{eq:gam-largeq} gives $0.844$ and $0.509$ at the same
point, 17\% and 29\% low.  These are $\epsilon \to 0$ roots; the seed
sits at $\epsilon = 0.18$, and the finite-$\epsilon$ roots the runs are
read against are computed in \ref{app:gam-closure} rather than
corrected by an imported factor.  ORB5 and GENE sit on the same relation at
$\epsilon = 0.1$, $q = 1.5$ and $k_r\rho_i = 0.055$ to within the 1\% their
published figure can be read to~\cite{Biancalani2017}.

\subsection{On a sheared profile the closed loop is a continuum in the field}
\label{app:gam-continuum}

The zonal test runs on the sheared profile $q(r) = q_0 + (q_a - q_0)(r/a)^2$,
so the local root varies across the seed and \eqref{eq:gam-disp} is not by
itself the reference.  Repeating the derivation with a general zonal
potential $\phi_0(r)$ in place of $e^{ik_r r}$ settles what the reference
is.  The drift operator is
$v_d(\sin\theta\,\partial_r + (\cos\theta/r)\,\partial_\theta)$, so the
sideband is driven by $\partial_r\phi_0$ at the particle's own radius, and
the $m = 0$ part of the return leg is
$(v_d/2i)(1/r)\partial_r(r h^+)$ and its conjugate --- a radial divergence
of the sideband's drift flux.  The zonal density is therefore
\begin{equation}
\delta n_0(r) = \frac{1}{r}\,\partial_r
  \left[\, r\, C(\omega; r)\, \partial_r\phi_0 \,\right] ,
\qquad
C(\omega; r) = -\frac{K_{\mathrm{full}}(\omega; q(r))}
                     {2\,\omega^2\Omega^2R^2} ,
\label{eq:gam-divergence}
\end{equation}
with $K_{\mathrm{full}}$ the whole right-hand side of
\eqref{eq:gam-disp}: the kernel sits \emph{between} the two radial
derivatives.  The zonal closure
$-(1/r)\partial_r(r\rho_s^2\partial_r\phi_0) = \tau\,\delta n_0$ then gives,
with $\tau C/\rho_s^2 = -K_{\mathrm{full}}/(2\omega^2R^2/v_t^2)$ from
$\rho_s^2 = \tau v_t^2/\Omega^2$,
\begin{equation}
\frac{1}{r}\,\partial_r
  \left\{\, r\,\rho_s^2\, \mathcal{R}(\omega; r)\, \partial_r\phi_0 \,\right\} = 0 ,
\qquad
\mathcal{R} = 1 - \frac{K_{\mathrm{full}}(\omega; q(r))}
                       {2\,\omega^2R^2/v_t^2} ,
\label{eq:gam-cont}
\end{equation}
whose solutions satisfy $r\,\mathcal{R}(\omega; r)\,E_r = \mathrm{const}$,
and regularity at the axis makes the constant zero.  The radial field can
be nonzero only where $\mathcal{R}(\omega; r) = 0$.  The spectrum is
therefore the continuum $\{\omega_G(r)\}$ of local roots, the
eigenfunctions are delta functions \emph{in the field}, and the
initial-value solution is
\begin{equation}
E_r(r,t) = E_r(r,0)\,e^{-i\omega_G(r)t} ,
\qquad
\phi(r,t) = \int_0^r g'(r')\,e^{-i\omega_G(r')t}\,\dd r' ,
\label{eq:gam-ivp}
\end{equation}
with $g$ the potential the seeded charge makes.  Every surface rings at its
own root in the field, and the zonal potential, which is what the run
reports, is the running integral of that winding field inside the surface.
The potential on a surface does not oscillate at that surface's own root.
At a flat $q$ the continuum is one frequency and the two statements
coincide, which is why the flat-$q$ leg of the test does not distinguish
them.

One consequence sets the reference the projected signal is read against.
The projection reads $P(t) = -\int W(r)\,g'(r)\,e^{-i\omega_G(r)t}\,\dd r$
after an integration by parts, with $W(r) = \int_0^r w\,(e - \langle e\rangle)$
the running integral of the mean-removed envelope weight.  The seeded charge
makes a $Q/r$ field outside the packet, the low-wavenumber part of the
inverse Laplacian.  Beyond the packet $W$ falls linearly to zero while $g'$
falls as $Q/r$, so the surfaces at $r/a = 0.6$ to $0.9$, whose roots are
$0.97$ down to $0.89\,c_s/a$, carry weight in the sum.  The fitter then
reads the frequency of a dephasing sum, which is not the envelope-weighted
mean of the roots it contains: at the test's point, built from the roots of
\eqref{eq:gam-disp}, the projected continuum reads $0.953$ where the
envelope-weighted root is $1.004$ and the root at the seed's own radius is
$1.018$.

\subsection{The torus's flux-surface average and the model's own root}
\label{app:gam-closure}

The derivation above takes the flux-surface average of the electron
response as the cylinder's, uniform in $\theta$, and lets
$\epsilon \to 0$.  The operator the code assembles is the torus's:
the projection of \paperEq{eq:fsa}{the flux-surface-averaged field
operator of the paper} carries the Jacobian $R = R_0(1 + \epsilon\cos\theta)$
in its weight (\paperSec{sec:method}{Section~2 of the paper}).  At the
seed's $\epsilon = 0.18$ the two averages differ at order $\epsilon$, and
the difference enters the closed loop through the one component the
electrons shield.  The $R$-weighted average of the $m = \pm1$ sideband is
not zero, so the sideband feeds the zonal electron response at order
$\epsilon$, a coupling the cylinder's average does not carry.  We have not
repeated the derivation of \ref{app:gam-disp} with the $R$-weighted average;
the model's root at finite $\epsilon$ is computed instead.

The computation is a linear initial-value solver for
\paperEq{eq:dk}{the drift-kinetic equation of the paper} on one
circular flux surface at a fixed radial wavenumber, independent of the
code (numpy only).  It keeps the mirror force and the $\theta$-dependence
of the drift, transports each orbit exactly in a Fourier series in the
orbit time, and closes the field with
\paperEq{eq:fsa}{the flux-surface-averaged field operator of the paper}
per $\theta$ node, with the average taken in either weight.  It is
radially local: the orbit's excursion does not feed back into $B$ or $q$,
the poloidal component of the magnetic drift is dropped (1\% to 3\% of the
rotation rate), and $k_\perp^2 = k_r^2$ in the polarization.  The root is
read from the zonal density over $[3, 60]\,a/c_s$ at the seed's
$k_r = 12.6/a$ on fifteen surfaces from $r/a = 0.15$ to $0.85$, and held
flat outside that span.

Table~\ref{tab:gam-roots} gives the roots across the seed with the average
in the cylinder's weight (\emph{flat}) and the torus's (\emph{toroidal}),
beside the roots of \eqref{eq:gam-disp}.  The flat column is the check.  It
reproduces the analytic relation in frequency to within 0.6\% for
$r/a \le 0.5$ and 1.3\% at $0.7$, and in damping to within 0.8\%, 3.1\% and
8.4\% at $r/a = 0.30$, $0.40$ and $0.50$.  Beyond $q = 1.6$ the two damping
columns part: the analytic root's damping is exponentially small there, and
the solver's finite $k_r$, which the relation does not carry, holds the
solver's damping at $0.0048$ and $0.0026\,c_s/a$ at $r/a = 0.60$ and
$0.70$, against the analytic $0.0037$ and $0.0005$.

\begin{table}[htbp]
\centering
\caption{Local GAM roots of the model at each flux surface, in $c_s/a$.
\emph{Analytic} is \eqref{eq:gam-disp}, the $\epsilon \to 0$ limit with the
cylinder's flux-surface average, at $k_r \to 0$.  \emph{Flat} and
\emph{toroidal} are the same model solved as an initial-value problem at the
seed's $k_r = 12.6/a$ with the cylinder's and the torus's average in the
electron closure; the latter is what the code assembles.  The analytic and
flat columns agree to the finite $k_r$ the relation does not carry.  The last
column is the closure's damping enhancement.}
\label{tab:gam-roots}
\small
\begin{tabular}{cc cc cc cc c}
\toprule
 & & \multicolumn{2}{c}{analytic} & \multicolumn{2}{c}{flat} & \multicolumn{2}{c}{toroidal} & \\
\cmidrule(lr){3-4} \cmidrule(lr){5-6} \cmidrule(lr){7-8}
$r/a$ & $q$ & $\omega$ & $\gamma$ & $\omega$ & $\gamma$ & $\omega$ & $\gamma$ & $\gamma_{\mathrm{tor}}/\gamma_{\mathrm{flat}}$ \\
\midrule
0.30 & 1.051 & 1.1495 & 0.0625 & 1.1429 & 0.0630 & 1.1375 & 0.0662 & 1.05 \\
0.40 & 1.203 & 1.0807 & 0.0339 & 1.0768 & 0.0350 & 1.0670 & 0.0405 & 1.16 \\
0.50 & 1.400 & 1.0181 & 0.0139 & 1.0125 & 0.0151 & 0.9982 & 0.0236 & 1.56 \\
0.60 & 1.640 & 0.9673 & 0.0037 & 0.9598 & 0.0048 & 0.9411 & 0.0169 & 3.52 \\
0.70 & 1.924 & 0.9303 & 0.0005 & 0.9187 & 0.0026 & 0.8982 & 0.0177 & 6.81 \\
\bottomrule
\end{tabular}
\end{table}

The torus's average lowers the frequency monotonically outward: by nothing
at $r/a = 0.15$, $1.4\%$ at the seed's centre and $2.6\%$ at $0.85$.  It
raises the damping by a factor that grows through the seed to a maximum
where the flat root's damping is least: $1.05$ at $r/a = 0.30$, $1.56$ at
the seed's centre, $6.8$ at $0.70$ and $3.2$ at $0.85$.  The two closures
part because the flat root's damping is the transit resonance, which falls
with $q$ as the resonant $\vpar = qR\omega$ moves into the tail of the
Maxwellian, from $0.109\,c_s/a$ at $r/a = 0.15$ to $0.003$ at $0.70$.  The
toroidal root holds $0.017$ to $0.019$ across the whole outer half,
$r/a = 0.60$ to $0.85$: the order-$\epsilon$ path through the electron
closure damps without that resonance, and it is the finite-aspect-ratio
damping enhancement Gao derives~\cite{Gao2010}.  The table is specific to
the seed: $k_r\rho_p$ runs from $1.05$ at $r/a = 0.15$ to $0.48$ at $0.60$,
and another $k_r$ gives another table.

\subsection{The runs read against the continuum}
\label{app:gam-reading}

We built the solution \eqref{eq:gam-ivp} from the local roots and the
seed's own potential, on the run's own radial bins, envelope weight,
surfaces and time grid, and read it through the run's own projection and
the same fit over $[0.05, 50]\,a/c_s$.  Two readings follow.
Tables~\ref{tab:gam-seeds} and~\ref{tab:gam-local} check the construction.
Their runs carry the electron closure in the cylinder's average, as the
continuum built from the roots of \eqref{eq:gam-disp} does, so run and
reference share a closure and differ only in what the construction leaves
out.  Table~\ref{tab:gam-matched} then reads the run in the code's own
closure against the continuum built from Table~\ref{tab:gam-roots}.

Table~\ref{tab:gam-seeds} compares run and continuum for every seed,
profile and shear the closed-loop test measured.  The run is $0.989$ of the
continuum at the test's point and $0.993$ at a flat $q = 1.4$.  The
continuum reproduces each of the run's dependences: the fall to $0.92$ at
$k_r = 6.3/a$, the width dependence, and the fall of the run's shortfall
below the local root from $7.5\%$ to $4.5\%$ ($0.942$ and $0.972$ against
$1.018$) when the magnetic shear is halved at fixed $q(r_0)$.  The two agree
to within 0.4\% to 1.2\% in every case but one, the $k_r = 25/a$ seed at
2.5\%, whose seed spans four cells of its mesh per wavelength.

\begin{table}[htbp]
\centering
\caption{The projected zonal signal, run against continuum, in $c_s/a$,
both with the electron closure in the cylinder's flux-surface average.
The continuum is \eqref{eq:gam-ivp} from the roots of \eqref{eq:gam-disp}
through the run's own projection and
fitter.  The runs are at $\Delta t = 0.1\,a/c_s$ on the previous step, whose
frequency is within 1.1\% of the corrected one's.  The first four rows
and the last are on the concentric
flux-aligned mesh; the $k_r = 25/a$ and half-shear rows are on the
unstructured disk mesh.}
\label{tab:gam-seeds}
\small
\begin{tabular}{lccc}
\toprule
Case & continuum $\omega$ & run $\omega$ & run / continuum \\
\midrule
profile, $k_r = 12.6/a$, $\sigma_e = 0.15$ & 0.9525 & 0.942 & 0.989 \\
profile, $\sigma_e = 0.3$                  & 0.9475 & 0.944 & 0.996 \\
profile, $\sigma_e = 0.075$                & 0.9480 & 0.936 & 0.988 \\
profile, $k_r = 6.3/a$, $\sigma_e = 0.3$   & 0.9263 & 0.922 & 0.995 \\
profile, $k_r = 25/a$, $\sigma_e = 0.075$  & 0.976\phantom{0} & 0.952 & 0.975 \\
half shear, $q_0 = 1.127$, $q_a = 2.219$ & 0.980\phantom{0} & 0.972 & 0.992 \\
flat $q = 1.4$                           & 1.0167 & 1.009 & 0.993 \\
\bottomrule
\end{tabular}
\end{table}

The projected signal is one number, and the continuum makes a local
prediction that is stronger: the field between neighbouring flux surfaces,
$-(\phi_{k+1}-\phi_k)/(r_{k+1}-r_k)$, oscillates at $\omega_G(r)$ on every
interval.  Table~\ref{tab:gam-local} reads that observable through a cubic
detrend, a Hilbert transform and an unwrapped phase, for the run and for
the continuum, over $[10,45]\,a/c_s$.  The run's field follows the local
root from $1.089$ at $r/a = 0.41$ to $0.898\,c_s/a$ at $0.91$, within
$0.3\%$ to $3.6\%$ of $\omega_G(r)$ and within $0.3\%$ to $3.0\%$ of the
continuum read through the same pipeline.  Nothing in the field has one
frequency.  The three intervals inboard of the seed Landau-damp at
$\gamma = 0.03$ to $0.06$ and carry no signal past $t = 30\,a/c_s$; over
$[5,20]$ they read $1.222$, $1.162$ and $1.107$ against the continuum's
$1.207$, $1.163$ and $1.119$, ratios of $1.012$, $0.999$ and $0.989$.

\begin{table}[htbp]
\centering
\caption{The zonal field between neighbouring flux surfaces, in $c_s/a$,
fitted over $[10,45]\,a/c_s$ at the test's point, run and continuum both
in the cylinder's closure.  $\omega_G(r)$ is the
local root of \eqref{eq:gam-disp} at $q_{\mathrm{eff}}(r)$.  The interval
at $r/a = 0.47$ is a node of the pattern in both and is not read; the three
intervals inboard of $0.41$ are in the text.}
\label{tab:gam-local}
\begin{tabular}{ccccc}
\toprule
$r/a$ & $\omega_G(r)$ & run & continuum & run / continuum \\
\midrule
0.406 & 1.076 & 1.089 & 1.077 & 1.011 \\
0.533 & 1.000 & 0.964 & 0.994 & 0.970 \\
0.597 & 0.969 & 0.949 & 0.966 & 0.982 \\
0.657 & 0.945 & 0.933 & 0.945 & 0.988 \\
0.717 & 0.925 & 0.920 & 0.925 & 0.995 \\
0.781 & 0.909 & 0.912 & 0.909 & 1.003 \\
0.844 & 0.897 & 0.907 & 0.898 & 1.011 \\
0.906 & 0.888 & 0.898 & 0.889 & 1.011 \\
\bottomrule
\end{tabular}
\end{table}

Table~\ref{tab:gam-matched} is the comparison in the code's own closure.
The seed of the first row of Table~\ref{tab:gam-seeds} runs on the
concentric mesh one level coarser than the campaign's at
$\Delta t = 0.1\,a/c_s$: at two wedges with the electron closure in each
average, at the campaign's $\alpha = 1.6\times10^{-2}$, and at eight wedges
in the torus's, at $\alpha = 10^{-3}$.  Each run is read against the
continuum \eqref{eq:gam-ivp} built from the matching column of
Table~\ref{tab:gam-roots}, through the run's own projection and fitter over
$[0.05, t_{\max}]$.  Matched at two wedges, the run in the torus's closure
is $0.4\%$ below its continuum in frequency at $t_{\max} = 50\,a/c_s$ and
$0.6\%$ at $100$ and $150$, and its damping is $0.82$, $0.77$ and $0.78$ of
the continuum's at the three windows.  The run in the cylinder's closure
reads $1.3\%$ to $1.6\%$ below the flat continuum in frequency and $0.72$ to
$0.80$ of it in damping.  At eight wedges the frequency is $0.4\%$ and
$0.2\%$ above the continuum and the damping $0.88$ and $0.94$ of it.  The
last row is the mismatch: the two-wedge torus-closure run against the
continuum of \eqref{eq:gam-disp}, the analytic column, reads $0.969$ of it
in frequency and $1.59$ times it in damping.  That is what the code's
default run gives against the reference of \ref{app:gam-disp}, and the
closure accounts for all of it.

\begin{table}[htbp]
\centering
\caption{The projected zonal signal against the continuum with run and
reference in the same electron closure, in $c_s/a$: L4,
$\Delta t = 0.1\,a/c_s$, the seed of the first row of
Table~\ref{tab:gam-seeds} at $\alpha = 1.6\times10^{-2}$ for the two-wedge
rows and $10^{-3}$ for the eight-wedge rows, fitted over
$[0.05, t_{\max}]$.  The continuum
is \eqref{eq:gam-ivp} from the named column of Table~\ref{tab:gam-roots}
through each run's own projection.
The toroidal rows are the code's default.  The two-wedge rows are on the
split-kick variant of \paperSec{sec:rh}{the zonal section of the paper}'s
configuration and the eight-wedge rows on the step
\paperAlg{alg:step}{the paper's step algorithm} lists.  The last
row reads the two-wedge toroidal run against the $\epsilon \to 0$
continuum, the mismatch that presents as over-damping.}
\label{tab:gam-matched}
\footnotesize
\setlength{\tabcolsep}{3pt}
\begin{tabular}{llcc cc cc cc}
\toprule
 & & & & \multicolumn{2}{c}{continuum} & \multicolumn{2}{c}{run} & \multicolumn{2}{c}{run / continuum} \\
\cmidrule(lr){5-6} \cmidrule(lr){7-8} \cmidrule(lr){9-10}
run closure & continuum & wedges & $t_{\max}$ & $\omega$ & $\gamma$ & $\omega$ & $\gamma$ & $\omega$ & $\gamma$ \\
\midrule
toroidal & toroidal & 2 & 50  & 0.9268 & 0.0377 & 0.9232 & 0.0310 & 0.996 & 0.82 \\
         &          &   & 100 & 0.9267 & 0.0407 & 0.9215 & 0.0313 & 0.994 & 0.77 \\
         &          &   & 150 & 0.9266 & 0.0408 & 0.9213 & 0.0319 & 0.994 & 0.78 \\
toroidal & toroidal & 8 & 50  & 0.9268 & 0.0377 & 0.9303 & 0.0331 & 1.004 & 0.88 \\
         &          &   & 100 & 0.9267 & 0.0407 & 0.9286 & 0.0383 & 1.002 & 0.94 \\
flat     & flat     & 2 & 50  & 0.9443 & 0.0231 & 0.9320 & 0.0185 & 0.987 & 0.80 \\
         &          &   & 100 & 0.9433 & 0.0282 & 0.9288 & 0.0204 & 0.985 & 0.72 \\
         &          &   & 150 & 0.9429 & 0.0290 & 0.9275 & 0.0211 & 0.984 & 0.73 \\
toroidal & analytic & 2 & 50  & 0.9525 & 0.0195 & 0.9232 & 0.0310 & 0.969 & 1.59 \\
\bottomrule
\end{tabular}
\end{table}

The seed amplitude does move the damping.  Table~\ref{tab:gam-alpha} is a
ladder at two wedges over four seeds spanning a factor of $160$.  The three
smallest agree in damping to $1\%$ and in frequency to $0.1\%$, and only
the campaign's $\alpha = 1.6\times10^{-2}$ departs, by $9\%$ in damping
and $0.6\%$ in frequency: the seed leaving the linear regime for this
observable.  The toroidal grid does not remove the departure: at the
campaign's seed eight wedges reads $0.73$ of the continuum and sixteen
wedges on the next finer mesh $0.74$, against two wedges' $0.82$.

\begin{table}[htbp]
\centering
\caption{The seed-amplitude ladder at two wedges, L4,
$\Delta t = 0.1\,a/c_s$, the torus's closure, fitted over
$[0.05, t_{\max}]$ against the toroidal continuum of
Table~\ref{tab:gam-matched}.  The last column is the residual of
\paperSec{sec:residual}{the residual section of the paper} on the same
runs, over the last quarter of $T = 150\,a/c_s$.  The damping and the
frequency are flat over the first three seeds and break at the fourth;
the residual does the opposite, because an unseeded run makes a
late-time signal of its own that enters the residual in proportion to
$1/\alpha$.  The campaign uses $\alpha = 1.6\times10^{-2}$ for the
residual and the linear limit for the damping.}
\label{tab:gam-alpha}
\begin{tabular}{lcccccc}
\toprule
 & \multicolumn{2}{c}{$t_{\max} = 50$} & \multicolumn{2}{c}{$t_{\max} = 100$} & \multicolumn{2}{c}{} \\
\cmidrule(lr){2-3} \cmidrule(lr){4-5}
$\alpha$ & $\omega$ & $\gamma$ & $\omega$ & $\gamma$ & $\gamma/\gamma_c$, $t_{\max} = 50$ & residual \\
\midrule
$10^{-4}$             & 0.9287 & 0.0338 & 0.9271 & 0.0383 & 0.90 & 0.187 \\
$10^{-3}$             & 0.9292 & 0.0342 & 0.9276 & 0.0378 & 0.91 & 0.061 \\
$4\times10^{-3}$      & 0.9283 & 0.0341 & 0.9262 & 0.0373 & 0.91 & 0.049 \\
$1.6\times10^{-2}$    & 0.9232 & 0.0310 & 0.9215 & 0.0313 & 0.82 & 0.044 \\
\bottomrule
\end{tabular}
\end{table}

The last column is why the two observables are measured at different
seeds.  At $\alpha = 0$ the same configuration makes a late-time
projection of $6.7\times10^{-3}$, a discretization floor that does not
scale with the seed and so enters the residual in proportion to
$1/\alpha$: $2\%$ of the residual at the campaign's $1.6\times10^{-2}$
and $86\%$ at $10^{-4}$.  Subtracting it leaves $0.0435$, $0.0452$ and
$0.0454$ at the three largest seeds, flat to $4\%$.  The damping is
fitted from the oscillation with a plateau constant of its own that
absorbs the floor.  So the residual needs the largest seed at which the
response is linear, the campaign's $1.6\times10^{-2}$, and the damping a
seed below it.

\subsection{Tools}
\label{app:gam-tools}

\begin{sloppypar}
Three scripts in the repository carry this appendix:
\texttt{scripts/\allowbreak ex11\_gam\_dispersion.py}, which solves
\eqref{eq:gam-disp} with a rational approximation of the Faddeeva
function checked against direct quadrature to $10^{-15}$ in both half
planes; \texttt{scripts/\allowbreak ex11\_zonal\_local\_ivp.py}, the initial-value
solver of \ref{app:gam-closure}, whose \texttt{-{}-fsa-weight}
selects the average and whose gate asserts the flat root at
$r/a = 0.5$ against \eqref{eq:gam-disp}; and
\texttt{scripts/\allowbreak ex11\_gam\_continuum.py}, which builds
\eqref{eq:gam-ivp} on a run's own bins and surfaces from either the
analytic roots or a table of computed ones (\texttt{-{}-roots}) and
reads both through the run's fitter.  The last carries a
\texttt{-{}-gate} that
asserts four properties of the construction, among them that at a flat
$q$ the continuum returns the local root through that fitter to $0.3\%$,
and that at the sheared profile its field reads $\omega_G(r)$ on every
interval to $1.5\%$ while its potential does not.
\end{sloppypar}
Table~\ref{tab:gam-roots} is generated by
\texttt{scripts/\allowbreak gen\_gam\_roots\_table.py} from the two root tables under
\texttt{paper/kintora-methods/data/}.

\section{The neoclassical polarization of the model}
\label{app:residual}

The zonal test of Section~\ref{sec:rh} compares a measured residual to the
collisionless neoclassical polarization of Xiao and
Catto~\cite{XiaoCatto2006}, an expansion derived for guiding-centre orbits in a
large-aspect-ratio tokamak.  \ref{app:gam} faced the same question for
the geodesic-acoustic frequency and answered it by re-deriving the dispersion
relation for Equation~(\ref{eq:dk}).  This appendix does the same for the
residual.  It shows that the flow of Section~\ref{sec:method} is an exact
guiding-centre flow, derives the residual for that flow, and evaluates it by
quadrature over the model's own orbits.  The evaluation reproduces the
published expansion to 2\% at zero wavenumber at the test's
$\epsilon = 0.18$ and to 0.1\% at the seed's wavenumber, so the reference is
the model's own and the comparison of Section~\ref{sec:rh} is sound up to
its radial locality (\ref{app:residual-observable}).

\subsection{The flow is an exact guiding-centre flow}
\label{app:residual-model}

Four properties establish this, and each is a statement about the circular
equilibrium of Section~\ref{sec:method} rather than about the discretization.

\emph{The rigid rotation is the true poloidal motion.}  For a concentric
circular equilibrium with poloidal flux $\psi(r)$ and $F = RB_\varphi$ constant,
the poloidal field $B_\theta = \psi'/R$ and the field strength
$|B| = \sqrt{F^2 + \psi'^2}/R$ carry the same factor $1/R$, so
\begin{equation}
\mathbf{b}\cdot\nabla\theta
 = \frac{B_\theta}{r|B|}
 = \frac{\psi'}{r\sqrt{F^2 + \psi'^2}}
\label{eq:res-bgradtheta}
\end{equation}
is constant on the surface.  The poloidal angle therefore advances uniformly in
time per unit $\vpar$, which is what the plane maps of
Section~\ref{sec:method} do.  With the code's $\psi' = B_0R_0\,r\iota\,g$ and
$F = B_0R_0$ the rate is $\iota g$ to within $0.4\%$ at $r_0/a = 0.5$.  What the
rigid line approximates is the \emph{toroidal} advance,
$\dd\varphi/\dd t = \vpar F/(R|B|)$, which the model replaces by a constant.  No
axisymmetric field depends on $\varphi$, so the zonal problem does not see that
approximation.  Xiao and Catto's own bounce average,
$\dd\tau = qR_0\,\dd\theta/\vpar$, is uniform in $\theta$ in the same
way~\cite{Xiao2006thesis}.

\emph{The invariant is the textbook one.}  The quantity
$p_\varphi = \psi + (F/|B|)\vpar$ is Littlejohn's canonical momentum for any
axisymmetric field, and the flow conserves it exactly.  The radial excursion at
fixed $p_\varphi$ is
\begin{equation}
\begin{gathered}
\delta(\theta)
 = -\frac{1}{\iota g\, r}
   \left[\frac{\vpar}{B} - \left\langle\frac{\vpar}{B}\right\rangle_{\!o}\right]
 = -\sigma\, v\, \rho_p\, s\, D(\theta) , \\
D = \frac{\sqrt{1 - \lambda b}}{b} - \left\langle\frac{\sqrt{1-\lambda b}}{b}\right\rangle_{\!o},
\end{gathered}
\label{eq:res-delta}
\end{equation}
with $\langle\cdot\rangle_o$ the orbit-time average, $b = B/B_0$,
$\lambda = \mu B_0/E$ the pitch, $\rho_p = q\rho_i/\epsilon$ the poloidal
gyroradius and $s = 1/(R_0g)$ the line-length pitch factor.  The excursion
differs from the physical one by the factor $s = 1.008$, which is $1.6\%$ of
$\Theta$.  We checked Equation~(\ref{eq:res-delta}) against a Runge--Kutta
integration of the guiding-centre equations themselves rather than assuming it:
the excursion agrees to $3.1\times10^{-3}$ in root mean square at
$\rho^* = 5\times10^{-4}$ and to $1.5\times10^{-3}$ at half that, first order in
$\rho^*$ as the finite-orbit-width remainder must be.  The trapped excursion
peaks at $0.395\,\rho_p$ and the passing one at $0.201\,\rho_p$.

\emph{The measure is the Liouville measure.}  The phase-space volume
element is $B^*_\parallel R\,\dd R\,\dd Z\,\dd\vpar\,\dd\mu$, and $B^*_\parallel \simeq B$
with $BR = B_0R_0$ constant, so the flat measure the stages carry is the correct
one.  The compressibility of the E$\times$B flow and the parallel dilation of the
curvature term cancel in it, which is the cancellation
Section~\ref{sec:rh} describes.

\emph{The energy exchange is complete.}  Along the flow,
$\dd(\vpar^2/2 + \mu B)/\dd t = -\vpar^2 D\,\partial_y\phi + \mu\,\mathbf{v}_E\cdot\nabla B$,
and the second term is $-\mu B D\,\partial_y\phi$, so the sum is
$-\mathbf{v}_d\cdot\nabla\phi$.  The curvature half is the acceleration sweep's
dilation and the grad-$B$ half is the fixed-$\mu$ E$\times$B advection across
field-strength contours.  Both $p_\varphi$ and $\vpar^2/2 + \mu B + \phi$ are
conserved, which is why the run without the curvature term does not settle at
all.

What remains approximate is $B^*_\parallel \to B$, the omission of $B_\theta^2$
from $|B|$ (0.4\% and constant on a surface), the plane Laplacian in the field
operator, and the plane projector in place of the toroidal flux-surface average.
The last two are the only ones that enter $\Theta$, and
\ref{app:residual-eval} measures the second at 0.7\%.

\subsection{The residual of the model}
\label{app:residual-theory}

Linearize Equation~(\ref{eq:dk}) about $f_0 = F_0(E)$ at flat profiles and write
$\delta f = -F_0\phi + h$.  The energy exchange above is the total time
derivative of $F_0(\vpar^2/2 + \mu B + \phi)$ along the unperturbed orbits, so
\begin{equation}
\frac{Dh}{Dt} = F_0\,\frac{\partial\phi}{\partial t} ,
\label{eq:res-h}
\end{equation}
with $D/Dt$ the transport along those orbits.  Integrating
Equation~(\ref{eq:res-h}) and letting the velocity moment phase-mix gives, in the
linear static-orbit theory,
\begin{equation}
\begin{gathered}
\frac{\phi(\infty)}{\phi(0)} = \frac{\mathcal{O}(k)}{1 + q^2\Theta/\epsilon^2} ,
\qquad
\mathcal{O}(k) = \Big\langle\!\Big\langle
  \big\langle e^{-ik\delta}\big\rangle_p\,
  \big\langle e^{ik\delta}\big\rangle_o
\Big\rangle\!\Big\rangle_p , \\
\Theta(k) = \frac{s^2\,[\,1 - \mathcal{O}(k)\,]}{c_B\,(k\rho_p s)^2} ,
\end{gathered}
\label{eq:res-residual}
\end{equation}
where $\langle\cdot\rangle_p$ is the average over the orbit with the plane
projector's weight, $\langle\!\langle\cdot\rangle\!\rangle_p$ adds the Maxwellian
average, and $c_B = \langle B_0^2/B^2\rangle$ is $1 + \epsilon^2/2$ for the plane
projector and $1 + 3\epsilon^2/2$ for the toroidal average.  Expanding
Equation~(\ref{eq:res-residual}) in $k$, the odd terms cancel between the two
signs of $\vpar$ and
$\Theta(0) = (3s^2/c_B)\langle\!\langle(\langle D^2\rangle_p + \langle D^2\rangle_o)/2\rangle\!\rangle_p$.
This is the form of Monreal et al.~\cite{Monreal2016} with the gyroaverage set to
unity, which is the drift-kinetic limit the model takes.

The published expansion does not carry the factor $\mathcal{O}(k)$.  Xiao and Catto impose an
initial charge through the classical polarization alone and do not transport it, so their residual is the ratio of
polarizations~\cite{Xiao2006thesis}.  The initial condition of
Section~\ref{sec:rh} multiplies $f_0$ by $1 + \alpha e(r)$, which the orbits do
carry, so the seed phase-mixes by its own orbit average.  At the test's point
$\mathcal{O} = 0.967$, and the single-surface number is therefore $0.1054$
rather than $0.1089$; the measured observable's reference, $0.085$, is in
\ref{app:residual-observable}.

\subsection{Evaluation}
\label{app:residual-eval}

We evaluate Equation~(\ref{eq:res-residual}) by quadrature: the trapezoid rule in
$\theta$ on passing orbits and the turning-point substitution of
Section~\ref{sec:ladder-bounce} on trapped ones, composite Gauss--Legendre in the
pitch graded geometrically into the trapped-passing boundary from both sides, and
Gauss--Hermite in the speed.  Four gates hold the evaluation: the closed-form
excursion against the guiding-centre equations
(\ref{app:residual-model}), the Rosenbluth--Hinton limit
$\Theta/\epsilon^{3/2} \to 1.633$ against the exact
$1.635$~\cite{Xiao2006thesis}, resolution independence to $10^{-6}$ relative
under a doubling of every quadrature, and the ranges of the residual and of
$\mathcal{O}$.

Table~\ref{tab:res-theta} compares the model's $\Theta$ with the published
expansion at zero wavenumber.  The two agree to 1\% below $\epsilon = 0.10$ and
depart by 6.5\% at $\epsilon = 0.30$, which is the $O(\epsilon^{7/2})$ remainder
the expansion carries.  Choosing the toroidal average over the plane one moves
$\Theta$ by 0.7\% at the test's $\epsilon$, so the projector is not a material
difference.  At the seed's wavenumber, $k_\perp\rho_p = 0.49$, the model gives
$\Theta = 0.1351$ against the expansion's $0.1353$, and a residual of $0.1054$
once the factor $\mathcal{O}$ is applied.

\begin{table}[htbp]
\centering
\caption{The neoclassical polarization $\Theta$ at $k_\perp = 0$: the model of
Section~\ref{sec:method} by quadrature over its own orbits, against the
expansion of~\cite{XiaoCatto2006}.  The two flux-surface averages differ by
0.7\% at the test's $\epsilon = 0.18$.}
\label{tab:res-theta}
\begin{tabular}{lcccc}
\toprule
$\epsilon$ & model, plane & model, toroidal & expansion & model / expansion \\
\midrule
0.05 & 0.01955 & 0.01974 & 0.01934 & 1.011 \\
0.10 & 0.05683 & 0.05758 & 0.05673 & 1.002 \\
0.15 & 0.10629 & 0.10746 & 0.10734 & 0.990 \\
0.18 & 0.14066 & 0.14166 & 0.14334 & 0.981 \\
0.25 & 0.23178 & 0.22977 & 0.24250 & 0.956 \\
0.30 & 0.30442 & 0.29686 & 0.32565 & 0.935 \\
\bottomrule
\end{tabular}
\end{table}

\subsection{The finite-wavelength coefficient of the reference}
\label{app:residual-audit}

The two published forms of the finite-wavelength correction disagree: the
thesis gives $-(k_\perp\rho_p)^2[\,(5/3)\epsilon^2 + 1.3\epsilon^{5/2} -
\ldots\,]$ and the article $-2.44\,\epsilon^{5/2}(k_\perp\rho_p)^2$ with no
term in $\epsilon^2$~\cite{Xiao2006thesis,XiaoCatto2006}.  A term in
$\epsilon^2$ cannot appear: writing the drift phase of a passing orbit as
$Q = \langle Q\rangle(1 + \tilde q)$ with $\langle\tilde q\rangle = 0$, the
fourth-order combination of the expansion collapses,
\begin{equation}
\tfrac{1}{12}\langle Q^4\rangle
+ \tfrac{1}{4}\langle Q^2\rangle^2
- \tfrac{1}{3}\langle Q\rangle\langle Q^3\rangle
= \langle Q\rangle^4\left[\tfrac{1}{12}\langle\tilde q^4\rangle
   + \tfrac{1}{4}\langle\tilde q^2\rangle^2\right] ,
\label{eq:res-cancel}
\end{equation}
so the bulk of the passing region enters at $O(\epsilon^4)$.  Evaluating
the reference's own integrals by the quadrature above, calibrated on its
wavenumber-independent bracket to $1.0009 \pm 0.0005$, the
wavenumber-dependent bracket divided by $\epsilon^{5/2}$ extrapolates to
$1.32$: the thesis's $\epsilon^{5/2}$ coefficient is right and its
$\epsilon^2$ term spurious.  The forms differ by 0.3\% in the residual at
the test's wavenumber, so the comparison of Section~\ref{sec:rh} does not
turn on the choice.

\subsection{The reference for the projected observable}
\label{app:residual-observable}

Equation~(\ref{eq:res-residual}) is the residual of one flux surface: the
settled-to-initial ratio of a potential whose radial shape the polarization
preserves.  The seed of Section~\ref{sec:rh} is not such a potential, for two
reasons.  Its density $\alpha\,e(r)$ has a net charge, $6.3\%$ of $\alpha$
over the disk, which the solver's removal of the mean charge turns into a
whole-disk response, so the seed's potential is the local $k_r = 12.6/a$
structure plus a smooth part that also projects onto the envelope.  And the
enhancement $1 + q^2\Theta/\epsilon^2$ that shields it rises from $7$ at
$r/a = 0.3$ through $10$ at the seed's centre to $32$ at the edge.  The
settled potential of the model is therefore the solution of the
polarization equation with that enhancement inside the divergence,
\begin{equation}
  -\frac{1}{r}\frac{d}{dr}\Big[\, r\,\rho_s^2\,\Big(1 + \frac{q^2\Theta(\epsilon)}{\epsilon^2}\Big)\frac{d\phi}{dr}\Big]
  = \tau\,\alpha\,\mathcal{O}\,\big(e(r) - \langle e\rangle\big),
  \label{eq:res-observable}
\end{equation}
with $\epsilon = r/R_0$, $q(r)$ the profile, $\rho_s^2$ averaged over the
surface with the plane measure ($c_B = 1 + \epsilon^2/2$), the natural
boundary condition and regularity on the axis.  We solve
Equation~(\ref{eq:res-observable}) and its initial counterpart (the
enhancement and $\mathcal{O}$ set to one) on a fine radial grid and project
both exactly as the run's binned potential is projected, on the run's own
bins, weights and envelope.  The initial solve reproduces the run's
$P(t_1)$ on the campaign mesh to $0.3\%$ ($7.05$ against $7.03$), which
checks the classical polarization of the discrete operator and the
flux-surface projector together.

Table~\ref{tab:res-observable} gives the settled ratio: $0.085$ through the
projected potential and through the radial field between neighbouring
surfaces over $0.35 < r/a < 0.70$, the two within $0.4\%$.  That is $18\%$
below the table's single-surface column at the seed's centre in every row
($19\%$ below the model's own single-surface $0.1054$), because the smooth
part of the potential and the outer half of the seed sit where the
enhancement is larger.  The model's settled field is not a scaled copy of
the seed's: $E_\infty/E_0$ is $0.125$ at $r/a = 0.4$, $0.093$ at $0.55$ and
$0.066$ at $0.7$, so the inner lobe of the seed is the part the model
shields least.  Equation~(\ref{eq:res-observable}) is the model's
$t \to \infty$ limit; Section~\ref{sec:residual} measures the model's
approach to it on the inner surfaces with the local initial-value solver.
It is also radially local: the enhancement at each radius acts on the
potential at that radius, so the equation neglects the orbits' width
against the radial scales of the seed and of $q$ and $\epsilon$, which
Section~\ref{sec:residual} weighs as a candidate for the measured gap.

\begin{table}[htbp]
\centering
\caption{The model's residual for the observable of Section~\ref{sec:rh},
from Equation~(\ref{eq:res-observable}) projected as the run is projected,
against the single-surface formula at $r_0$.  $\mathcal{O}$ is the seed's
phase-mixing factor of Equation~(\ref{eq:res-residual}); the last row
scales the expansion's $\Theta$ to the model's finite-wavelength value
($0.1351/0.14066$ at the test's point).}
\label{tab:res-observable}
\small
\setlength{\tabcolsep}{4pt}
\begin{tabular}{lccc}
\toprule
$\Theta$, $\mathcal{O}$ & single surface at $r_0$ & $P(\infty)/P(t_1)$ & field ratio \\
\midrule
expansion at $k \to 0$, $\mathcal{O} = 1$     & 0.1034 & 0.0849 & 0.0847 \\
expansion at $k \to 0$, $\mathcal{O} = 0.967$ & 0.1000 & 0.0821 & 0.0819 \\
expansion's $\Theta$ scaled, $\mathcal{O} = 0.967$ & 0.1036 & 0.0852 & 0.0849 \\
\bottomrule
\end{tabular}
\end{table}

\subsection{Tools}
\label{app:residual-tools}

\begin{sloppypar}
Three scripts in the repository carry this appendix.
\texttt{scripts/\allowbreak ex11\_model\_residual.py} evaluates the
polarization: its \texttt{-{}-gate} asserts the four properties above,
\texttt{-{}-scan} evaluates the residual at the radii of the test, and
\texttt{-{}-xc} performs the audit of \ref{app:residual-audit}.
\texttt{scripts/\allowbreak ex11\_model\_observable.py} solves
Equation~(\ref{eq:res-observable}) on a run's own bins with that tool's
equilibrium and $\Theta$, and prints Table~\ref{tab:res-observable}.  The
expansion itself is evaluated by
\texttt{scripts/\allowbreak ex11\_rh\_residual.py}, whose gates hold its
validity range.
\end{sloppypar}

\newif\ifanon \anonfalse
\section{Artifact description}
\label{app:artifact}

\begin{sloppypar}
\textbf{What the artifact is.}  \ifanon This description serves this paper
and its companion paper. \else This description serves two papers:
\emph{Parallel design of a GPU-resident semi-Lagrangian drift-kinetic solver
on unstructured extruded tokamak meshes} (the HPC paper) and \emph{KINTORA: a
semi-Lagrangian discontinuous Galerkin drift-kinetic, unstructured mesh
tokamak solver} (the methods paper).\fi{}  Their claims rest on one code base,
one gate ladder and one set of measurement files.  The code base is the
drift-kinetic solver, the \texttt{kin} library under \texttt{src/kin} and its
driver \texttt{ex11}, built on PETSc and Kokkos, with the 1X+3V
Vlasov--Poisson solver (\texttt{ex6}) and the 2X+2V constant-advection driver
(\texttt{ex8}) that an extended build of the HPC paper's source reports.  \ifanon The artifact
is a public repository, withheld here for double-anonymous review, \else The
artifact is the \texttt{kintora-hpc} and \texttt{kintora-methods} directories
of \url{https://gitlab.com/markadams4/petsc-claude-papers}, \fi which carry the
papers, the solver source frozen with the methods paper, the measurement data
and the scripts that reduce and plot it; each directory's \texttt{README}
records the commit.  Both papers' runs ran on earlier builds of the same tree:
the HPC paper's binaries are identified below by their checksums, and the
methods paper's data \texttt{README}s name the binary and source commit of
each run.  The reduced files the papers' numbers are
read from are in the artifact; the raw job outputs they were reduced from are
held in the run directories on the machine.

\textbf{Software.}  PETSc, \ifanon the authors' \else the author's \fi fork of
it, configured with the Kokkos and CUDA (\texttt{sm\_80}) backends; the fork is
needed for its device-side finite-element assembly and, to a lesser extent, its
device-side multigrid setup.  The GPU logs of both papers that carry PETSc's
revision line record the same commit of it, \texttt{80c04f7} of 2026-09-17, on
its main branch, and the workstation gates build against the same
fork\ifanon{}
(the fork is withheld here for double-anonymous review)\else{}
(\texttt{petsc-claude}; its current tree is the \texttt{petsc/} directory of
\url{https://gitlab.com/markadams4/petsc-claude-papers}, whose
\texttt{petsc/FORK-COMMITS.txt} lists the fork's commits, \texttt{80c04f7}
among them)\fi.
Kokkos as configured by PETSc; GPU-aware Cray MPICH; a GCC host compiler.
\texttt{src/kin/makefile} builds the library from its host and device units,
and the drivers' makefile links it.  Every option is listed under
\texttt{-help}.

\textbf{Hardware.}  Perlmutter (NERSC) GPU nodes: one AMD EPYC 7763 and four
NVIDIA A100s (40 and 80\,GB) per node on Slingshot-11, one MPI rank per GPU.
Perlmutter CPU nodes (two EPYC 7763) for the rank-invariance sweeps.
Development gates run on Apple M-series hardware with the Kokkos Serial
backend, which is also where the CPU half of every CPU-versus-GPU reproduction
is measured.

\textbf{Gates.}  The acceptance ladder is a committed script with committed
reference values, and it is what admits a change to the solver:
\texttt{scripts/check\_ex11\_vadvance.sh} on a workstation, and
\texttt{scripts/campaign/submit\_gate\_split.sh} on the machine, five
dependent jobs on the debug queue that take 82 minutes on one GPU node between
them, the longest 28.  Each physics gate asserts a measured rate against a
reference computed independently of the code, and the tools that compute those
references assert their own known degenerations before they are used.
\texttt{scripts/bytecheck\_ex11.sh} compares the physics output of two binary
sets byte for byte on 51 configurations; it is the arbiter of every commit
that moves code and changes no arithmetic.

\textbf{Experiment workflow.}  From the repository root,
\texttt{scripts/campaign/deploy.sh} copies the sources, meshes and scripts to
the machine by \texttt{rsync}, verifies every file by its checksum, and ships a
git bundle from which the gate ladder clones its historical rungs.  The series
binaries are built in a tree of their own and copied to the run directory, so a
build never races a gate.  Each timed run reports the PETSc
\texttt{-log\_view} events of its hot steps: a point of the strong and weak
series is 105 steps, of which the first five warm up in their own log stage,
except at the third weak-series point, which runs 30, and the ragged A/B runs
100 steps after one.  The runs use \texttt{-log\_view\_gpu\_time}.

\textbf{The HPC paper's measurements.}  The geometry of every timed run is
\texttt{-pol\_mesh\_file meshes/solovev\_L4.mesh -pol\_dm\_refine 0
-eq\_shape solovev}, with \texttt{solovev\_L5} and \texttt{solovev\_L6} at the
second and third weak-series points, and \texttt{-pol\_curved} off.  The
options the configuration turns on are \texttt{-nrow\_blocks} (the $\mu$ row
axis), \texttt{-field\_row\_split} (the per-plane solves divided among the row
ranks), \texttt{-field\_fsa} with \texttt{-fsa\_degree 2} (the electron
response), \texttt{-exb} (the E$\times$B term as a semi-Lagrangian stage),
\texttt{-phi\_pc\_type cholesky -phi\_pc\_factor\_mat\_solver\_type cudss}
(the direct per-plane solve), \texttt{-field\_batch\_stations} and
\texttt{-tor\_halo\_device\_mpi}, the last with
\texttt{MPICH\_GPU\_SUPPORT\_ENABLED=1} and \texttt{-{}-gpu-bind=none}.  The
one-sided worst-case exchange is the same binary with
\texttt{-tor\_halo\_worst\_case}, bit-identical in its physics to the ragged
one.  Each series has a comma-separated file under \texttt{data/} whose rows
carry the job they came from:
\begin{itemize}
\item the strong series over $\mu$ row blocks: \texttt{rowshaped\_submit.sh}
  over \texttt{ws\_x.sbatch}, jobs 58794292, 58794296, 58794298, 58794300 and
  58794302 with the solves divided, and 58804473, 58804475, 58804476 and
  58804477 with every row rank solving, a second allocation on a later binary;
  \texttt{row\_series.csv};
\item the accuracy-matched weak series: \texttt{ws16.sbatch} at
  \texttt{PT=A} and \texttt{PT=B}, jobs 58779096 and 58779097, with the
  field-stage entries from the device-timed reruns 58981406 and 58981408, and
  \texttt{ws16Cfix.sbatch}, job 58789831, at the third point;
  \texttt{weak\_series\_16x.csv};
\item what each optimization is worth: \texttt{ws16Cabl.sbatch}, job 58794289,
  with \texttt{ws16Bexb.sbatch}, job 58794291, for the E$\times$B row at the
  second point and \texttt{ws16Crep.sbatch}, job 58804472, for the
  divided-solve case; \texttt{ablation\_c.csv};
\item the roofline of the third point's rank: \texttt{rfC.sbatch}, job
  58877367, with the L2 and L1 columns from a second Nsight Compute run of the
  same binary, job 59180182; \texttt{roofline\_pointC.csv};
\item the ragged exchange in time: \texttt{raggedab.sbatch}, job 58867319, on
  the binary job 58867317 built; \texttt{ragged\_ab.csv};
\item the shaped plane's cost against the circular one at matched sizes:
  matched pairs of \texttt{ws16.sbatch}, jobs 58770209 to 58770212, at the
  first two weak-series points, and the circular control in the third
  point's allocation, job 58789831; \texttt{shaped\_vs\_circular\_stages.csv}
  and \texttt{weak\_series\_16x.csv};
\item the rank shapes tried at the third weak-series point: jobs 58778611
  and 58789831; \texttt{point\_c\_shapes.csv};
\item the per-plane direct solve against plane size, and the tuned algebraic
  multigrid it replaced: job 58704153 on one A100-80GB;
  \texttt{solve\_vs\_plane\_size.csv};
\item the cost of the flux-surface-averaged electron response: the
  \texttt{fsacost} line every weak-series run prints;
\item the velocity-grid comparison, uniform against equal-mass $v_\parallel$
  segments: \texttt{vg4.sbatch}, job 59241072, on the circular plane;
  \texttt{velocity\_grid\_vg4.csv};
\item the rank- and node-invariance sweep on CPU nodes:
  \texttt{cpu\_np\_sweep.sbatch}, on the solver of 2026-08-31;
  \texttt{rank\_invariance.csv};
\item the run-to-run repeats behind the timing spread quoted below:
  \texttt{repeats.csv};
\item for the extended build only, the plane-count series
  (\texttt{ws\_x.sbatch} at \texttt{SERIES=plane}, job 58170363;
  \texttt{plane\_series.csv}) and the A/B of the semi-Lagrangian E$\times$B
  stage against the Eulerian source on the circular plane (jobs 58165464
  and 58170789; \texttt{exb\_ab.csv}).
\end{itemize}
The binaries are \texttt{ex11\_ws19} (md5 \texttt{c3ec463a}), \texttt{ex11\_ws21}
(\texttt{ba45f3cf}), \texttt{ex11\_ws23} (\texttt{872086a1}) and
\texttt{ex11\_ws24} (\texttt{96dbebe5}), each built by its own job under
\texttt{scripts/campaign/}.  The GPU series take 27 node-hours together, and
the third weak-series point carries almost all of it: a run there holds 64
nodes, which makes the weak series $6.5$ node-hours and the ablation $20$,
against $0.44$ for the strong series in both of its solve modes.  Bit-identical
reproduction is expected for the rank sweep and for the row axis at one block;
the divided solves reproduce the every-rank ones to $2.9\times10^{-15}$, which
is the direct solver across devices; timing reproduces within the run-to-run
spread of a shared machine, measured at $3.0\%$ and $3.3\%$ of the mean on
repeated points.  Every point of the series is a single run in its own batch
allocation, so a difference below about $5\%$ between two single runs is not
an effect.

\textbf{The methods paper's figures and tables.}  Every figure has a committed
script that reads a committed data file and prints the numbers the text quotes:
\texttt{gen\_mms\_figure.py}, \texttt{gen\_bounce\_figure.py},
\texttt{gen\_orbit\_figure.py}, \texttt{gen\_landau\_figure.py},
\texttt{gen\_floor\_figure.py}, \texttt{gen\_coherence\_figure.py} and
\texttt{gen\_itg\_budget\_figure.py}, each named for its figure.  None is a
pixel-identical clone of the committed image, and two of them locate their
reference features, the envelope fit window and the recurrence time, from the
trace itself; each script's docstring says how.  The conservation and mass
tables are printed by \texttt{gen\_conservation\_tables.py} from the run logs,
and the table of local geodesic-acoustic roots by
\texttt{gen\_gam\_roots\_table.py}; the other appendix tables are typeset in
their \texttt{.tex} files, from the tools \texttt{scripts/ex11\_gam\_*.py},
\texttt{scripts/ex11\_zonal\_local\_ivp.py},
\texttt{scripts/ex11\_model\_residual.py} and
\texttt{scripts/ex11\_model\_observable.py}.  The zonal-flow logs those
tools read are in \texttt{data/zonal/}, gzipped extracts (the binned
potential and density, the surface averages and the run banners) of every
job the zonal tables cite, with a \texttt{README} that maps each file to its
row; the fitters read the gzipped files directly.  The toroidal ITG rates are the
legs of \texttt{scripts/campaign/cbc6.sbatch}, one A100 each, with a
\texttt{README} in \texttt{data/} that maps every leg to its job; the
demonstration is job 59271321, 1,200 steps in 532\,s with its halo staged
through the host; its repeat with the device-direct halo, job 59483229,
reproduces it to the printed digits in 413\,s, 332\,ms per step, and is the
timing the papers quote.

\textbf{The zonal runs.}  The zonal-flow measurements of the methods paper run
on GPU nodes through \texttt{scripts/campaign/rhtor.sbatch} and the runners
beside it that sweep the toroidal wedge count (\texttt{rhtorN.sbatch}, and
\texttt{rhcpuN.sbatch} on CPU nodes), run the default step as a control
(\texttt{rhtor\_def.sbatch}) and pair a circular control with the shaped plane
(\texttt{rh\_shaped8.sbatch}, which fixes both); the others set the seed
amplitude and the electron closure through their environment.
Every one of those runs is identified by its wedge count, its seed amplitude,
its closure and its step variant, and no number from them can be read without
all four: the closure moves the geodesic-acoustic damping by a factor of $1.5$
to $1.7$, by fit window, two wedges read the residual 38\% to 44\% below eight,
and the seed amplitude moves the damping by $9\%$ and the unsubtracted residual
by a factor of four.  The torus's flux-surface average in the electron closure
is the code's default; a driver built before that default must be given the
option explicitly.

\textbf{Known gaps.}  The right panel of the slab-ITG budget
figure, a four-configuration control set, has no data file; its left panel is
regenerated in full.  Two figures built earlier in the work are committed but
referenced by neither paper, and one of them cannot be regenerated because the
options of its run were not recorded.  The rank-invariance sweep's job
outputs were kept without their job numbers, so its file holds the values
recorded from them when the sweep was committed.  The commit and line counts
quoted for the restructuring of the monolithic solver into the library come
from the development history, which the artifact does not carry; the artifact
carries the code at the commit its \texttt{README} names.  The batch scripts carry the project
account and absolute paths of the machine they ran on, and must be edited
before they run elsewhere.
\end{sloppypar}

\bibliographystyle{elsarticle-num}
\bibliography{references}

\end{document}